\documentclass[%
reprint,
superscriptaddress,
 amsmath,amssymb,
aps,
prb,
]{revtex4-2}

\usepackage{graphicx}
\usepackage{dcolumn}
\usepackage{bm}
\usepackage{hyperref}
\usepackage{amsthm}
\usepackage{comment}
\usepackage[capitalise]{cleveref}
\hypersetup{
    colorlinks = true,
    allcolors=black,
    citecolor=blue,
    linkcolor=blue
} 
\usepackage{amsmath, amsfonts, dsfont, mathrsfs, float,amssymb}
\usepackage{color}
\usepackage[caption=false]{subfig}
\usepackage{svg}
\usepackage{tikz}
\usepackage{tikz-network}
\usepackage{orcidlink}
\usepackage{xcolor}
\usepackage{mathtools}
\usepackage{multirow}
\usepackage{csquotes}
\usepackage{svg}

\newtheorem{theorem}{Theorem}

\begin{document}
\makeatletter

\title{A Random Matrix Theory of Pauli Tomography} 
\author{Nathan Keenan}
\email{nakeenan@tcd.ie}
\affiliation{School of Physics, Trinity College Dublin, Dublin 2, Ireland}
\affiliation{Trinity Quantum Alliance, Unit 16, Trinity Technology and Enterprise Centre, Pearse Street, D02 YN67, Dublin 2, Ireland}
\affiliation{IBM Quantum, IBM Research Europe - Dublin, IBM Technology Campus, Dublin 15, Ireland}

\author{John Goold}
\email{gooldj@tcd.ie}
\affiliation{School of Physics, Trinity College Dublin, Dublin 2, Ireland}
\affiliation{Trinity Quantum Alliance, Unit 16, Trinity Technology and Enterprise Centre, Pearse Street, D02 YN67, Dublin 2, Ireland}
\affiliation{Algorithmiq Ltd, Kanavakatu 3C 00160, Helsinki, Finland}
\author{Alex Nico-Katz}
\email{nicokata@tcd.ie}
\affiliation{School of Physics, Trinity College Dublin, Dublin 2, Ireland}
\affiliation{Trinity Quantum Alliance, Unit 16, Trinity Technology and Enterprise Centre, Pearse Street, D02 YN67, Dublin 2, Ireland}
\date{\today}
\begin{abstract}

Quantum state tomography (QST), the process of reconstructing some unknown quantum state $\hat\rho$ from repeated measurements on copies of said state, is a foundationally important task in the context of quantum computation and simulation. For this reason, a detailed characterization of the error $\Delta\hat\rho = \hat\rho-\hat\rho^\prime$ in a QST reconstruction $\hat\rho^\prime$ is of clear importance to quantum theory and experiment. In this work, we develop a fully random matrix theory (RMT) treatment of state tomography in informationally-complete bases; and in doing so we reveal deep connections between QST errors $\Delta\hat\rho$ and the gaussian unitary ensemble (GUE). By exploiting this connection we prove that wide classes of functions of the spectrum of $\Delta\hat\rho$ can be evaluated by substituting samples of an appropriate GUE for realizations of $\Delta\hat\rho$. This powerful and flexible result enables simple analytic treatments of the mean value and variance of the error as quantified by the trace distance $\|\Delta\hat\rho\|_\mathrm{Tr}$ (which we validate numerically for common tomographic protocols), allows us to derive a bound on the QST sample complexity, and subsequently demonstrate that said bound doesn't change under the most widely-used rephysicalization procedure. These results collectively demonstrate the flexibility, strength, and broad applicability of our approach; and lays the foundation for broader studies of RMT treatments of QST in the future.


\end{abstract}

\maketitle

\section{Introduction}
\label{sec:intro}

Quantum state tomography (QST) \cite{bagan2004collective, cai2016optimal, keyl2006quantum, guctua2008optimal} is the process of experimentally reconstructing an unknown quantum state $\hat\rho$ from the outcomes of repeated measurements on copies of said state. Efficient and precise implementation of QST is a foundational task in quantum computation and simulation: state verification, experimental determination of non-trivial functions of $\hat{\rho}$ \cite{nico2024can}, and channel tomography \cite{yu2023almost, huang2024learning} (via e.g. the Choi-Jami\l kowski isomorphism), all hinge on high-fidelity QST. Moreover, QST bears importance in the context of quantum foundations: the ontological state $\hat{\rho}$ is fundamentally inaccessible to any `classical' process, and hence there exist limitations on how accurately (implicit) phenomenal tomographic reconstructions can reflect reality. Thus, the behaviour (and bounding) of errors in QST, here quantified by the trace distance $\|\hat\rho-\hat\rho^\prime\|_\mathrm{Tr} = \|\hat\Delta\hat\rho\|_\mathrm{Tr}$, are of basic importance across theoretical, experimental, and computational quantum physics. In this work, we address this problem using a complete random matrix theory treatment of tomographic processes. In doing so we reveal deep connections between states produced by tomographic protocols and the gaussian unitary ensemble (GUE), and develop and deploy a powerful method for evaluating functions over empirical spectral measures.

In this work, we are solely concerned with complete, single-copy, fixed shot count QST: in which we seek to reconstruct $\hat\rho$ as completely as possible given access to exactly one copy of $\rho$ for any single measurement \cite{flammia2012quantum, lowe2022lower}. This represents the (widely used) prototypical form of QST deployed in e.g. native QST on IBM devices. Examples of other QST protocols include those that allow simultaneous measurements on multiple entangled copies of $\rho$ \cite{yuen2023improved, bubeck2020entanglement, o2016efficient, o2017efficient, haah2016sample, chen2024optimal, flammia2024quantum}, protocols that partially reconstruct the state such as shadow tomography \cite{akhtar2209scalable, zhang2024quantification, sengupta2025partial, rath2021quantum, aaronson2018shadow, elben2023randomized, chen2022exponential, chen2024optimal, huang2020predicting} and quantum overlapping tomography \cite{garcia2020pairwise, cotler2020quantum}, and state verification of $\hat\rho^\prime$ against known \cite{buadescu2019quantum, chen2022tight, liu2024role, guctua2020fast} or unknown \cite{yu2023almost} ontological quantum states $\hat\rho$. For a more comprehensive overview of possible tomographic protocols (and related topics), we refer the reader to Ref.~\cite{acharya2025pauli}. 

\begin{figure}
    \centering
        \includegraphics[width=\linewidth]{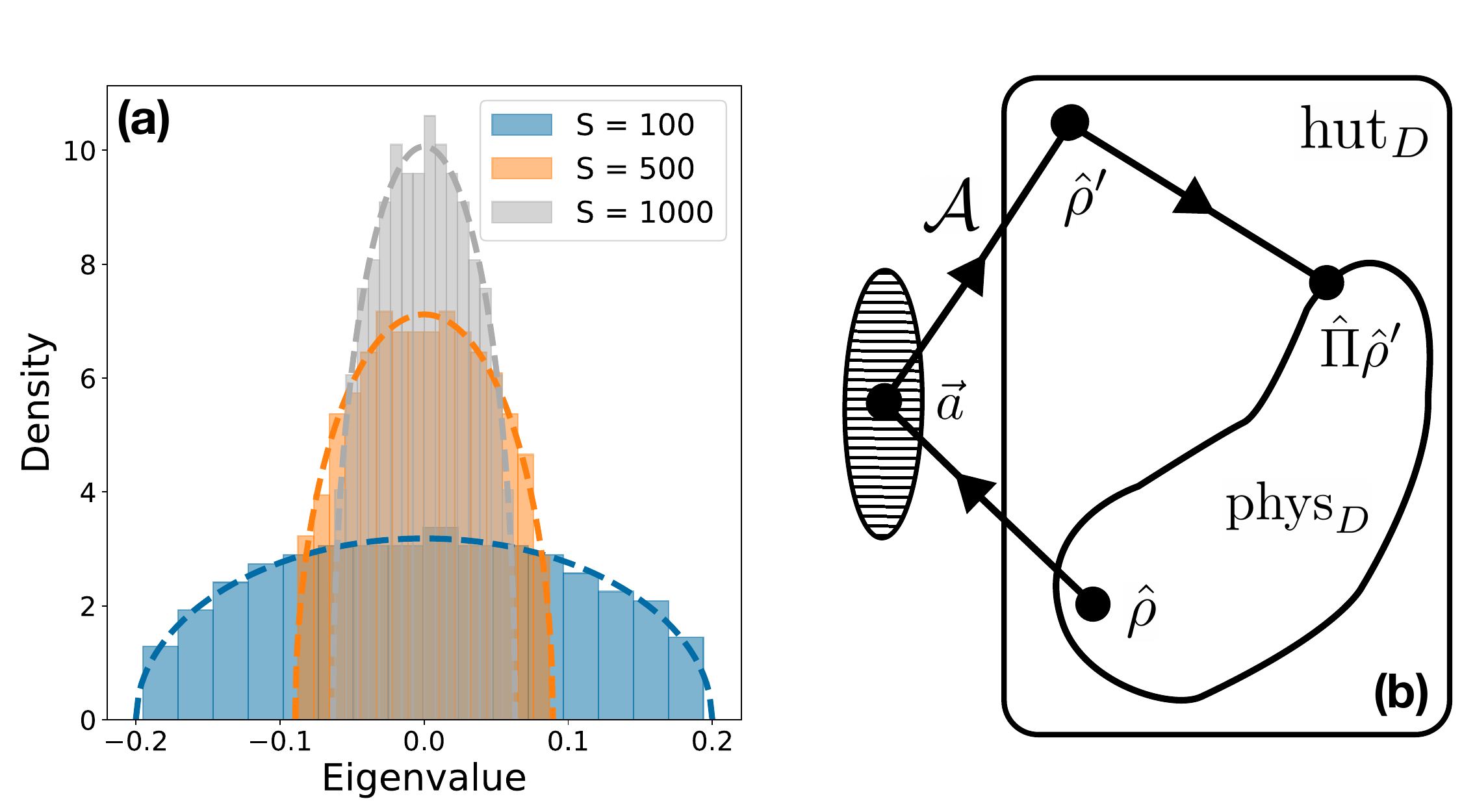}
    \caption{\textbf{(a)} Eigenvalue distributions of naive tomographic excess $\Delta\hat\rho$ for different shot counts $S$, overlayed with the GUE predicted Wigner semi-circles of radius $2/\sqrt{S}$. \textbf{(b)} Schematic for tomographic procedure. An a priori state is measured in an IC-basis, which leads to coefficients $\vec a$. The basis expansion map $\mathcal{A}(\vec a)$ then gives us an estimate $\hat\rho'$ of $\hat\rho$. If we then want to ensure that our estimator for $\hat\rho$ is physical, we project back to $\mathrm{phys}_D$ via $\Pi$ which depends on the rephysicalization scheme.}
    \label{fig:schematic}
\end{figure}

We first introduce and motivate the basic elements of the full single-copy QST protocols we consider in \cref{sec:motivation}. In \cref{sec:gue}, we introduce the GUE measure and develop a series of theorems which connect it to full QST in a fixed (symmetric) informationally-complete basis; this forms the basis of our full RMT treatment of tomographic errors. In \cref{sec:likely} we develop two more theorems which bound the difference between Lipschitz functions over the one-point and two-point empirical spectral measures of the tomographic excess $\Delta\hat\rho$, and identical functions evaluated over a (rescaled) GUE spectral measure; we then demonstrate that these distances vanish for both a na\"ive QST protocol (in \cref{sec:naive}) and a widely used more sophisticated QST protocol (in \cref{sec:sophisticated}). This is the major result of our work: that for common tomographic protocols, one can substitute random samples from the GUE for the random error matrix $\Delta\hat\rho$ in calculations of functions of the QST error. 

Equipped with this powerful and generic result, we proceed to: calculate analytic values for the expected value $\mathbb{E}[\|\Delta\rho\|_\mathrm{Tr}]$ and variance $\mathrm{var}[\|\Delta\rho\|_\mathrm{Tr}]$ of QST errors in \cref{sec:errors}, validate these derived values numerically in \cref{sec:numerics}, validate a recently calculated bound (c.f. Ref.~\cite{acharya2025pauli}) on the QST sample complexity in \cref{sec:complexity}, and finally demonstrate that the most widely-used process of state rephysicalization (i.e. artificial enforcement of positivity after determination of $\hat\rho$) does not improve the existing complexity bound in \cref{sec:rephysicalization}. In doing so, we demonstrate the strength and general flexibility of our RMT approach to QST, and conclude by speculating on further possible avenues of research in \cref{sec:speculate}.

\section{Motivation}
\label{sec:motivation}

Given an $N$-qubit system, an arbitrary state of the system $\rho \in \mathbb{C}^{D\times D}$ (where $D=2^N$ is the total dimension) can be written in the orthogonal Pauli basis as
\begin{equation}
\label{eq:pauli_basis}
    \rho = \frac{1}{D}\sum_{j} c_j \hat P_j,
\end{equation}
where $c_j = \mathrm{Tr}(\rho P_j)$, and $P_j$ runs over all the $N$-length Pauli strings $\mathcal{P}_N = \{\mathbb{I}, X, Y, Z\}^{\otimes N}$. Full state Pauli tomography consists of performing measurements in the basis defined by elements of $\mathcal{P}_N$ on copies of the state $\hat\rho$ \cite{cai2016optimal, gross2010quantum, acharya2025pauli}. Since the basis $\mathcal{P}_N$ is informationally-complete (IC) (i.e. spans the physical state space), an unbiased estimator $\hat\rho^\prime$ for $\hat\rho$ can then be constructed: 
\begin{equation}\label{eq:tomo-basis}
 \hat\rho^\prime = \frac{1}{D}\sum_{j} a_j \hat P_j
\end{equation}
where the $a_j$ are ultimately derived from the measurement data (see \cref{sec:naive} and \cref{sec:sophisticated} for a detailed discussion). As mentioned in \cref{sec:intro} we consider only fixed shot-count $S$ (non-adaptive) QST protocols (c.f. Ref.~\cite{mahler2013adaptive, chen2023does}); however we are nonetheless able to validate the recent bound of Ref.~\cite{acharya2025pauli} in \cref{sec:complexity}, supporting the conjecture made by the authors of Ref.~\cite{acharya2025pauli} that their bound is tight, at least in the non-adaptive case \footnote{Non-adaptive QST protocols are those in which the decision on which POVM element $P_j \in \mathcal{P}_N$ to measure during a given shot is independent of previous shot outcomes.}. One of the main issues with such a reconstruction is that \cref{eq:tomo-basis} may yield a density matrix $\hat\rho^\prime \in \mathrm{hut}_D$ which is hermitian and unit trace but is not positive semi-definite and thus is clearly aphysical $\hat\rho^\prime \notin \mathrm{phys}_D$. This is due to the fact that measurement outcomes are random and (for states which are not full-rank) \cref{eq:tomo-basis} induces a probability measure with non-trivial support on negative eigenvalues for arbitrary shot counts $S$; we remark on this issue from an RMT perspective in \cref{sec:complexity}. For a single qubit, $\mathrm{phys}_D$ maps onto the Bloch sphere, but for larger systems the geometry of $\mathrm{phys}_D$ is complicated and difficult to characterize \cite{dietz2006generalized, wie2020two, filatov2024towards}. The typical approach to dealing with this positivity issue is to project $\hat\rho^\prime$ back into $\mathrm{phys}_D$, where $\mathrm{phys}_D$ is a convex subset $\mathrm{phys}_D\subset \mathrm{hut}_D$ of $\mathrm{hut}_D$, via minimisation over a chosen norm \cite{singh2016constructing, smolin2012efficient, ferrie2018maximum, siddhu2019maximum}. In the particular case of the maximum likelihood estimator (c.f. Ref.~\cite{smolin2012efficient}), we later show in \cref{sec:rephysicalization} that this minimization does not improve the sample complexity compared to the non-rephysicalized tomographic state $\hat\rho^\prime$.

We finally remark that a central object of our study $\Delta\hat\rho$ can be expanded in terms of the coefficients of \cref{eq:pauli_basis} and \cref{eq:tomo-basis} as
\begin{equation}\label{eq:premap}
    \Delta\hat\rho = \sum_j (c_j-a_j) \hat P_j = \sum_j y_j \hat P_j
\end{equation}
where $y_j$ are random variable distributed around zero. Determining connections between the matrices induced by $\vec{y}$ via \cref{eq:premap} (formalized as \cref{eq:effect-map}) and the GUE ensemble forms the content of \cref{sec:gue} and \cref{sec:likely}.

\section{The GUE in the Pauli Basis}
\label{sec:gue}

In this section we are concerned with relationships between probability measures on subspaces of the space $\mathrm{Mat}_D(\mathbb{C})$ of $D\times D$ matrices over $\mathbb{C}$, and probability measures on $\mathbb{R}^{D^2}$. The central object of interest is the \textit{Gaussian Unitary Ensemble} (GUE), a probability measure on the $D^2$ dimensional real subspace of Hermitian matrices $\mathcal{H}_D\subset\mathrm{Mat}_D(\mathbb{C})$ with density given by independent Gaussians over its independent entries (i.e. those elements not fixed by enforcing hermiticity), each with total variance $\sigma^2$ \cite{mehta2004random}:
\begin{align}
\label{eq:gue}
\mathrm{d}\mu(H)&\propto \exp\left(
    -\sum_{j} \frac{H_{jj}^{2}}{2\sigma^2} - \sum_{j < k} \frac{[H_{jk}^{(r)}]^2}{\sigma^2} - \sum_{j<k}\frac{[H_{jk}^{(i)}]^2}{\sigma^2}
    \right)
 \mathrm{d}H \\
&= \exp\left(- \frac{1}{2\sigma^2}\mathrm{Tr}(H^2)\right)\mathrm{d}H,
\end{align}
where $H^{(r)}_{jk}$ and $H^{(i)}_{jk}$ are the real and imaginary components of $H_{jk}$ respectively, and $\mathrm{d}H$ is induced by the Lebesgue measure on $\mathbb{R}^{D^2}$:
\begin{equation}
    \mathrm{d}H = \prod_{j\leq k} \mathrm{d}H_{jk}^{(\mathrm{r})}\prod_{j<k}\mathrm{d}H_{jk}^{(\mathrm{i})}.
\end{equation}
It is worth mentioning that the interest in this measure comes from the spectra of the resultant sampled matrices. For example, the single eigenvalue probability distribution function for such matrices at leading order in $D$ is given by the \textit{Wigner semi-circle distribution}:
\begin{equation}\label{eq:wigner}
    W_R(\lambda) = \frac{2}{\pi R^2}\sqrt{R^2-\lambda^2},
\end{equation}
with radius $R=2\sigma\sqrt{D}$.

The main result of this section is \cref{tm:pauli-theorem}, which is a corollary of a more general statement \cref{tm:ic-theorem} about informationally complete measurements procedures. An informationally complete (IC) measurement procedure is given by an operator basis $\{E_j\}$ which spans $\mathcal{H}_D$. Given such a basis $\{E_j\}$, we can define the dual frame $\{F_j\}$ such that $\mathrm{Tr}(E_jF_k) = \delta_{jk}$. If also each of the $E_j$ are positive semidefinte, and they sum to the identity, we call $\{E_j\}$ an informationally complete POVM \cite{d2004informationally, garcia2021learning}. More generally though, an IC-basis naturally admits an expansion of operators in $\mathcal{H}_D$ as a weighted sum of the $\{E_j\}$ defined by the map
\begin{align}\label{eq:effect-map}
    \mathcal{A}_E&:\mathbb{R}^{D^2}\rightarrow \mathcal{H}_{D}\subset \mathrm{Mat}_{D}(\mathbb{C}):\vec{a} \rightarrow \sum_{j} a_j E_j,\\
    a_j &= \mathrm{Tr}(HF_j).
\end{align}
Since the basis $\{E_j\}$ spans $\mathcal{H}_D$, $\mathcal{A}_E$ is a linear isomorphism on its range, and so the inverse map $\mathcal{A}_E^{-1}$ is well defined. Finally, we introduce the Gaussian measure $\mathrm{d}\mu_\Sigma$ on $\mathbb{R}^{D^2}$:
\begin{equation}\label{eq:gaussian-measure}
d\mu_\Sigma(\vec{a}) \propto\exp\left(-\frac{1}{2}\vec{a}^\top\Sigma^{-1}\vec{a}\right)da
\end{equation}
where $\mathrm{d}a$ is the Lebesgue measure on $\mathbb{R}^{D^2}$. Here, $d\mu_\Sigma$ corresponds to sampling from the multivariate Gaussian distribution $\mathcal{N}(0, \Sigma)$ with means $0$ and covariance matrix $\Sigma$. Such a random vector $\vec{a}$ naturally arises in tomographic settings (see \cref{sec:motivation}); and the map \cref{eq:effect-map} can be understood as a tomographic reconstruction protocol given some experimentally realizable IC-basis.
\begin{theorem}\label{tm:ic-theorem}
    Given the Gaussian measure $\mathrm{d}\mu_\Sigma$ on $\mathbb{R}^{D^2}$ of \cref{eq:gaussian-measure} with diagonal covariance matrix $\Sigma = \mathrm{diag}(v_1,v_2,\cdots,v_{D^2})$; then $\mathcal{A}_E$ defined with respect to an IC-basis $\{E_j\}$ with dual frame $\{F_j\}$ induces the Gaussian measure
    \begin{equation}\label{eq:ic-measure}
        \mathrm{d}\mu_\Sigma^\mathrm{IC} \propto \exp\left[-Q(H)\right]\mathrm{d}H
    \end{equation}
    on $\mathcal{H}_D$, where $Q(H)$ is the positive-definite quadratic form given by:
    \begin{equation}\label{eq:quadratic-form}
        \frac{1}{2}\sum_j \frac{1}{v_j}\mathrm{Tr}(HF_j)^2.
    \end{equation}
\end{theorem}
\textit{Proof.} Pushing forward \cref{eq:gaussian-measure} by $\mathcal{A}_E$ induces the following measure $d\mu_\Sigma(H)$ on the space of real Hermitian matrices $\mathcal{H}_{D}$:
\begin{equation}\label{eq:proof-measure}
        d\mu_\Sigma(H) \propto\exp\left(-\frac{1}{2}[\mathcal{A}_E^{-1}(H)]^T\Sigma^{-1}\mathcal{A}_E^{-1}(H)\right)dH.
\end{equation} 
For $\Sigma = \mathrm{diag}(v_1,v_2,\cdots,v_{D^2})$, and using the inverse map $\vec{a} = \mathcal{A}_E^{-1}(H)$, the exponent of \cref{eq:proof-measure} reduces to
\begin{equation}\label{eq:ic}
    Q_\mathrm{IC}(H) = \frac{1}{2}[\mathcal{A}_E^{-1}(H)]^T\Sigma^{-1}\mathcal{A}_E^{-1}(H) = \frac{1}{2}\sum_j \frac{a_j^2}{v_j}
\end{equation}
where we can use the dual frame $\{F_j\}$ to retrieve the $a_j = \mathrm{Tr}(HF_j)$. Since $\{E_j\}$ is informationally complete, $Q(H)$ must be positive definite. \qed

We now prove two corollaries of \cref{tm:ic-theorem}. The first for IC orthogonal bases, where we have the additional property that $\mathrm{Tr}(E_jE_k) = C\delta_{jk}$ for some constant $C$; in this setting we retrieve the GUE of \cref{eq:gue} for uniform $v_j = v$. The second for symmetric informationally-complete (SIC) POVMs which satisfy additional constraints $E_j = \Pi_j/D$ and $\mathrm{Tr}(\Pi_j\Pi_k) = D(\delta_{jk}+1)/(D+1)$, and yields a quadratic form $Q$ which is additionally unitarily invariant. The former corollary forms the central result around which the rest of this article is structured; whilst a deeper study of the implications of both the more general result for IC-POVMs, and the more specific result for SIC-POVMs, we defer to future research.

First we turn our attention to the case of an orthogonal IC-basis, of which the set of Pauli strings $\{P_j\}$ is an example.
\begin{theorem}\label{tm:pauli-theorem}
     Given the Gaussian measure $\mathrm{d}\mu_\Sigma$ on $\mathbb{R}^{D^2}$ of \cref{eq:gaussian-measure} with uniform diagonal covariance matrix $\Sigma = v\mathbb{I{}}$; then $\mathcal{A}_E$ defined with respect an IC-basis $\{E_j\}$ satisfying $\mathrm{Tr}(E_j E_k) = C\delta_{jk}$ induces the GUE measure on $\mathcal{H}_D$.
\end{theorem}
\textit{Proof.} The coefficients $a_j$ resulting from the inverse map $\mathcal{A}_E^{-1}(H)$ can be retrieved as $a_j = \mathrm{Tr}(HE_j)/C$ with dual frame $F_j=E_j/C$. Now consider the expansion of $\mathrm{Tr}(H^2)$ in the $\{E_j\}$ basis:
\begin{equation}
    \mathrm{Tr}(H^2) = \sum_{jk} a_ja_k\mathrm{Tr}(E_j E_k) = C\sum_ja_j^2
\end{equation}
and thus that
\begin{equation}\label{eq:theorem-trace-rule}
    \sum_j \mathrm{Tr}(H_j P_j)^2 = \frac{1}{C}\mathrm{Tr}(H^2).
\end{equation}
Direct substitution of \cref{eq:theorem-trace-rule} into \cref{eq:quadratic-form}, and setting $v_j= v$ yields the quadratic form
\begin{equation}\label{eq:pauli-quadratic}
    Q_\textrm{P}(H) = -\frac{1}{2Cv}\mathrm{Tr}(H^2)
\end{equation}
which, after substitution into \cref{eq:ic-measure} yields the measure
\begin{equation}\label{eq:pauli-measure}
        \mathrm{d}\mu_\Sigma^P(H) \propto \exp\left(-\frac{1}{2Cv}\mathrm{Tr}(H^2)\right)dH,
\end{equation}
which is identical in form to the GUE measure of \cref{eq:gaussian-measure} with rescaled variances $\sigma^2 = vC$. \qed

Such an IC-basis is the Pauli basis, in which case $C=D=2^N$.
Due to these rescaled variances, the eigenvalue distribution induced by \cref{eq:pauli-measure} is given at leading order by the Wigner semi-circle distribution of \cref{eq:wigner} but with appropriately rescaled Wigner radius
\begin{equation}\label{eq:wigner-radius}
    R = 2D\sqrt{v}.
\end{equation}

Finally, given widespread recent interest in symmetric informationally-complete (IC) POVMS, one might consider how the induced GUE measure of \cref{eq:ic-measure} changes should we insist on a basis expansion in terms of such a SIC-POVM instead \cite{renes2004symmetric, stricker2022experimental, gour2014construction}.

\begin{theorem}\label{tm:sic-theorem}
     Given the Gaussian measure $\mathrm{d}\mu_\Sigma$ on $\mathbb{R}^{D^2}$ of \cref{eq:gaussian-measure} with uniform diagonal covariance matrix $\Sigma = v \mathbb{I}$; then $\mathcal{A}$ defined with respect to a SIC-POVM induces a unitarily invariant quadratic form $Q_\mathrm{SIC}(H)$.
\end{theorem}

\textit{Proof.} The proof proceeds identically to that of \cref{tm:pauli-theorem}; with the notable substitution of the SIC-POVM effect definition $E_j = \Pi_j/D$ where $\Pi_j$ is a projector, and the SIC-POVM overlap definition $\mathrm{Tr}(\Pi_j\Pi_k) = D(\delta_{jk}+1)/(D+1)$. Proceeding identically to the calculation of \cref{eq:pauli-quadratic}, we find the SIC-POVM quadratic form
\begin{equation}\label{eq:sicpovm}
    Q_\mathrm{SIC}(H) = \frac{D \mathrm{Tr}(H^2) + \mathrm{Tr}(H)^2}{D(D+1)},
\end{equation}
which is unitarily invariant. \qed

\section{Reconstruction error as a likely sampling of the GUE}
\label{sec:likely}

The basic full tomographic reconstruction process of a quantum state $\hat{\rho}$ starts with some experimentally realizable informationlly complete measurement procedure, and implements the basis expansion map $\mathcal{A}_E$ as per \cref{eq:effect-map} by making measurements to determine the $a_j$ experimentally. Throughout the rest of this work we restrict our discussion to the Pauli IC-basis $\{P_j\}$, wherein the $P_j$ all have eigenvalues $\pm 1$. Thus for a given string $P_j$ one can experimentally determine a list of measurement outcomes $X_{j} \in \{-1,1\}^{\times S_j}$ where $S_j$ is the shot count for Pauli string $P_j$ (i.e. the total number of experimental `shots' carried out). The mean value $x_{j} = \overline{X}_{j}$ of any such realization is clearly a binomially distributed random variable with mean $a_j = \mathrm{Tr}(\hat{\rho}P_j)/D$ and variance $v_{j} = (1-\mathrm{Tr}(\hat{\rho}P_j)^2)/(S_jD^2)$. One can then utilize the ``Pauli expansion map" $\mathcal{A}_P$ to reconstruct a tomographic copy $\hat{\rho}^\prime_{S} = \sum_{j} x_{j} P_j$ of the ontological (ultimately inaccessible) state $\hat{\rho}$. The basic principle of tomography is thus clearly demonstrated: taking each $S_j$ to sufficiently high values yields accurate enough statistics $x_{j} \approx a_j$ to enable a reconstruction of the ontological state $\hat \rho$ up to some tolerance $\varepsilon$, i.e, $\|\Delta\hat\rho\| = \|\hat{\rho}^\prime-\hat{\rho}\| \leq \varepsilon$ for some metric of interest $\|,\|$. In this work we will be interested in the trace norm $\|\Delta\hat\rho\|_\mathrm{Tr} = \sum_j|\lambda_j(\Delta\hat\rho)|$ which is the sum over the absolute value of the eigenvalues. It is useful at this point to note that we can write the matrix $\Delta\hat\rho = \hat\rho'-\hat\rho = \sum_jy_jP_j$, where $y_j = x_j - a_j$. Thus each $y_j$ is binomially distributed with mean $0$ and equal variance to $x_j$. 

There are two immediate problems with applying Theorem~\ref{tm:pauli-theorem} to the object $\Delta\hat\rho$. First of all, the $y_j$ are not Gaussian. For large shot counts $S_j$ we can invoke the central limit theorem, but more formally, \cite{tao2011random} contains results on the universality of certain eigenvalue statistics for random matrices with entries sampled from distributions with sub-Gaussian tails. In particular, see condition $C0$ and \textit{Theorem 15} therein. For the rest of the paper, we will thus assume that the coefficients $y_j$ in the Pauli expansion of $\Delta\hat\rho$ are Gaussian. Secondly, even with Gaussian $\vec y \sim \mathcal{N}(0, \Sigma)$, the covariance matrix $\Sigma$ will in general not be diagonal, let alone proportional to the identity. If each of the $4^N$ Pauli coefficients are calculated independently from seperate experiments, then $\Sigma$ will be diagonal. To illustrate our method, we will begin with this assumption, and then show how the procedure changes when a more efficient experimental setup is used. However, we still have the problem that Theorem~\ref{tm:pauli-theorem} assumes that $\Sigma$ is proportional to the identity. We will show in Theorem~\ref{tm:vanishing} that for what we are interested in calculating, we may replace $\Sigma =\mathrm{diag}(v_1,v_2,\cdots,v_{D^2}) $ with $\Sigma = \overline v\mathbb{I}$ where $\overline v = (1/D^2)\sum_j v_j$.

\begin{theorem}\label{tm:vanishing}\
A random Gaussian vector $\vec{y} \sim \mathcal{N}(0,\Sigma)$ with diagonal, non-uniform, covariance matrix $\Sigma = \mathrm{diag}(v_1,v_2\cdots,v_{D^2})$, and an IC-basis choice $\{E_j\}$ satisfying $\mathrm{Tr}(E_jE_k) = D\delta_{jk}$ yields a matrix $\mathcal{A}_E(\vec{y})$ with spectral measure $\mu_y(\lambda)$. Given a sampling of $\vec{y}$, one can induce a sampling of a GUE matrix $\mathcal{A}_E(\vec{g})$ with spectral measure $\mu_g(\lambda)$; such that the probability $\mathbb{P}$ that the difference between Lipschitz functions $f$ integrated over said measures exceeds $a$ is bounded by:
\begin{equation}\label{eq:tm4-res}
    \mathbb{P}\left[\left|\int f\mathrm{d}\mu_y - \int f\mathrm{d}\mu_g\right|^2 \geq a||f||_\mathrm{Lip}^2\right] \leq \frac{D}{a\overline{v}}\mathrm{var}[v],
\end{equation}
where $\overline{v} = \sum_j v_j / D^2$ and $\mathrm{var}[v] = \sum_j(v_j-\overline{v})^2/D^2$.
\end{theorem}
\textit{Proof.} The structure of this proof involves the introduction of a sequence of upper bounds for the l.h.s. of \cref{eq:tm4-res} of decreasing strictness; first in terms of the Wasserstein distances $W_\alpha(\mu_y,\mu_g)$ \cite{beatty2025wasserstein} between measures, then the Frobenius norm $\|\mathcal{A}_E(\vec{y})-\mathcal{A}_E(g)\|_F$, and finally (via Markov's inequality) by the r.h.s. of \cref{eq:tm4-res}.

We proceed directly by remarking that the l.h.s. of \cref{eq:tm4-res} is bounded by the Wasserstein-1 distance $W_1$ as follows:
\begin{equation}\label{eq:wass-1}
    \left|\int f\mathrm{d}\mu_y - \int f\mathrm{d}\mu_g\right|^2 \leq \|f\|_\mathrm{Lip}^2 W_1(\mu_y, \mu_g)^2
\end{equation}
where $\|f\|_\mathrm{Lip}$ is the Lipschitz norm of $f$. We here find it useful to explicitly define the induced measures:
\begin{equation}\label{eq:one-point}
    \mu_r(\lambda) = \frac{1}{D}\sum_j \delta\left(\lambda - \lambda_j^{(r)}\right)
\end{equation}
where $\lambda_j^{(r)} = \lambda_j(\mathcal{A}_E(\vec{r}))$ denotes the $j$-th (sorted) eigenvalue of the matrix $\mathcal{A}_E(\vec{r})$ induced by $\vec{r}$, and where $\delta$ is the Dirac delta function. The Wasserstein-1 distance is bounded by $W_1(\mu_y, \mu_g)^2 \leq W_2(\mu_y, \mu_g)^2$ where the Wasserstein-2 distance between a homogenous sum of $n$ atoms in $\mathbb{R}^d$ is given by 

\begin{equation}
    W_2\left(\frac{1}{n}\sum_j\delta_{\vec a_j}, \frac{1}{n}\sum_j\delta_{\vec b_j}\right)^2 = \underset{\sigma\in S_d}{\mathrm{min}}\frac{1}{d}\sum_j\|\vec a_j - \vec b_{\sigma(j)}\|^2_2
\end{equation}
and so since the eigenvalues are already sorted, we have:
\begin{equation}\label{eq:w2-dist}
    W_2(\mu_y, \mu_g)^2 = \frac{1}{D^2}\sum_j \left|\lambda_j^{(y)}-\lambda_j^{(g)}\right|^2.
\end{equation}

We here make use of the \textit{Hoffman-Wielandt inequality} which states that, for normal complex matrices $A, B \in \mathrm{Mat}_{d,d}(\mathbb{C})$ with eigenvalues $\{\lambda_j(A)\}$ and $\{\lambda_j(B)\}$ respectively, then:
\begin{equation}\label{eq:hw-inequality}
    \min_{\sigma \in S_d} \sum_{j=1}^d \left| \lambda_j(A) - \lambda_{\sigma(j)}(B) \right|^2 \leq \| A - B \|_F^2
\end{equation}
where $S_d$ is the permutation group of $\{1,2,\cdots,d\}$. As $\mathcal{A}_E(\vec{y})$ and $\mathcal{A}_E(\vec{g})$ are both Hermitian, and thus normal, we can apply the Hoffman-Wielandt inequality of \cref{eq:hw-inequality} directly to \cref{eq:w2-dist} to upper-bound the Wasserstein distance $W_2$ as follows:
\begin{equation}
    W_2(\mu_y, \mu_g)^2 \leq \frac{1}{D^2}\sum_j \|\mathcal{A}_E(\vec{y}) -\mathcal{A}_E(\vec{g})\|_F^2
\end{equation}
where we have used the fact that the default ascending ordering of eigenvalues solves the minimization problem over permutations. This result reduces \cref{eq:wass-1} to the more tractable form
\begin{equation}\label{eq:frob-1}
    \left|\int f\mathrm{d}\mu_y - \int f\mathrm{d}\mu_g\right|^2 \leq \frac{\|f\|_\mathrm{Lip}^2}{D^2} \|\mathcal{A}_E(\vec{y}) -\mathcal{A}_E(\vec{g})\|_F^2.
\end{equation}
We now remark that, given \cref{eq:frob-1}, the following inequality holds over the distribution of $\vec{y}$:
\begin{widetext}
\begin{equation}\label{eq:frob-prob}
    \mathbb{P} \left[ \left|\int f \mathrm{d}\mu_y - \int f\mathrm{d}\mu_g\right|^2 \geq a \|f\|_\mathrm{Lip}^2 \right] \leq \mathbb{P} \left[ \frac{1}{D^2}\|\mathcal{A}_E(\vec{y}) -\mathcal{A}_E(\vec{g})\|_F^2 \geq a \|f\|_\mathrm{Lip}^2 \right],
\end{equation}
\end{widetext}
and, as both sides of \cref{eq:frob-prob} are positive, we can invoke Markov's inequality $\mathbb{P}[X \geq a] \leq X/a$ to manipulate the r.h.s. of \cref{eq:frob-prob} into the following form:
\begin{equation}\label{eq:frob-2}
    \mathbb{P} \left[ \frac{1}{D^2}\|\mathcal{A}_E(\vec{y}) -\mathcal{A}_E(\vec{g})\|_F^2 \geq a \right] \leq \frac{\mathbb{E}\left[\|\mathcal{A}_E(\vec{y}) -\mathcal{A}_E(\vec{g})\|_F^2\right]}{ a D^2}.
\end{equation}

We now turn our attention to the evaluation of $\mathbb{E}[\|\mathcal{A}_E(\vec{y}) -\mathcal{A}_E(\vec{g})\|_F^2]$. We proceed by fixing a form for $\vec{g}$ based on $\vec{y}$ and computing the expectation value directly; first rewrite the elements of $\vec{y}$ as $y_j = z_j\sqrt{v_j}$ in terms of the dummy statistic $z_j \sim\mathcal{N}(0,1)$. We can then introduce a corresponding vector $\vec{g}$ of values $g_j = z_j\sqrt{\overline{v}}$, such that $(y_j-g_j)/z_j = \eta_j = \sqrt{v_j}-\sqrt{\overline{v}}$. By \cref{tm:pauli-theorem}, $\mathcal{A}_E(\vec g)$ is a GUE sampled matrix. Using \cref{eq:effect-map} we can explicitly construct the Frobenius norm as follows:
\begin{equation}
    \|\mathcal{A}_E(\vec y) - \mathcal{A}_E(\vec g)\|_F^2 = \left\|\sum_j\eta_j z_j E_j\right\|_F^2 =
     D\sum_j (\eta_j z_j)^2,
\end{equation}
where we have exploited the orthogonality relation $\mathrm{Tr}(E_iE_j)=D\delta_{ij}$. Taking the expectation value of both sides yields
\begin{equation}
    \mathbb{E}\left[\|\mathcal{A}_E(\vec y) - \mathcal{A}_E(\vec g)\|_F^2\right] = D\sum_{j}\eta_j^2
\end{equation}
where we have exploited the fact that $\mathbb{E}[z_j^2] = \mathrm{var}[z_j] = 1$ by definition. From the earlier definition of $\eta_j = \sqrt{v_j}-\sqrt{\overline{v}}$ we can determine the following inequality for $\eta_j^2$:
\begin{equation}
    \eta_j^2 = \frac{(v_j - \overline{v})^2}{\left(\sqrt{v_j}+\sqrt{\overline{v}}\right)^2} \leq \frac{(v_j - \overline{v})^2}{\overline{v}}
\end{equation}
such that the expectation value of the Frobenius norm is upper bounded by
\begin{equation}\label{eq:frob-bound}
    \mathbb{E}\left[\|\mathcal{A}_E(\vec y) - \mathcal{A}_E(\vec g)\|_F^2\right] \leq \frac{D^3}{\overline{v}}\mathrm{var}[v].
\end{equation}
Direct substitution of \cref{eq:frob-bound} into \cref{eq:frob-2} and \cref{eq:frob-prob} after rescaling $a$ by the factor $\|f\|_\mathrm{Lip}^2$; yields \cref{eq:tm4-res}, completing the proof. \qed

\begin{theorem}\label{tm:vanishing-twopoint}
    The condition \cref{eq:tm4-res} (rescaled by a factor of two) holds for integrals of Lipschitz functions over the empirical two-point spectral measures.
\end{theorem}
The proof is extremely similar to the above, and so we will simply sketch the differences. The empirical two-point spectral measure $\mu_r^{(2)}$ is defined as 
\begin{equation}\label{eq:two-point}
        \mu^{(2)}_r(\lambda, \lambda') = \frac{1}{D^2(D^2-1)}\sum_{j\neq k}\delta(\lambda-\lambda_j^{(r)})\delta(\lambda'-\lambda_k^{(r)})
\end{equation}
which is the empirical form of the integrand in Eq.~(6.1.2) in \cite{mehta2004random}. Then assuming ascending ordering of the eigenvalues as before:
\begin{widetext}
\begin{equation}
    W_2(\mu^{(2)}_y, \mu^{(2)}_g)^2 = \frac{1}{D^2(D^2-1)}\sum_{j\neq k}\left(\left|\lambda_j^{(y)} - \lambda_j^{(g)}\right|^2 + \left|\lambda_k^{(y)} - \lambda_k^{(g)}\right|^2\right) \sim \frac{2}{D^2}\sum_j\left|\lambda_j^{(y)} - \lambda_j^{(g)}\right|^2 = 2 W_2(\mu_y, \mu_g)^2,
\end{equation}
\end{widetext}
such that rest of the proof proceeds identically to \cref{tm:vanishing} (rescaled by a factor of two). \qed

\begin{theorem}\label{tm:vanishing-nondiag}
    \cref{tm:vanishing} and \cref{tm:vanishing-twopoint} can be extended to non-diagonal $\Sigma$ in the following sense. If $\vec y \sim \mathcal{N}(0, \Sigma)$, 
    then for Lipschitz functions $f$, \cref{eq:tm4-res} holds, where the $v_j$ are the eigenvalues of $\Sigma$.
\end{theorem}
Since $\Sigma$ is real and symmetric, $\Sigma = U\Lambda U^T$
for some orthogonal $U$, and $\Lambda = \mathrm{diag}(v_1,...,v_{D^2})$. Then $\vec y = U\vec z$ where $\vec z\sim \mathcal{N}(0, \Lambda)$. Also,
\begin{equation}
    \mathcal{A}_E(\vec y) = \mathcal{A}_E(U\vec z) = \sum_{jk}U_{jk}z_k E_j = \mathcal{A}_{E'}(\vec z),
\end{equation}
where $E'_k = \sum_jU_{jk}E_j$. Now all that is left to show is that $\mathrm{Tr}(E'_jE'_k) = D\delta_{jk}$, and so we can replace $\mathcal{A}_E(\vec y)$ in \cref{tm:vanishing} with $\mathcal{A}_{E'}(\vec z)$. 
\begin{align}
    &\mathrm{Tr}(E'_jE'_k) = \mathrm{Tr}\left[\left(\sum_\ell U_{\ell j} E_\ell\right)\left(\sum_{m}U_{mk} E_m\right)\right] \\ &= \sum_{\ell m}U_{\ell j}U_{mk}\mathrm{Tr}\left(E_\ell E_m\right) = D\sum_{\ell m}U_{\ell j} U_{mk} \delta_{\ell m} \\ &= D\sum_\ell U_{\ell j}U_{\ell k} = D(U^TU)_{jk} = D\delta_{jk}.
\end{align}
And so $\mathcal{A}_{E'}(\vec z) = \mathcal{A}_{E}(\vec y)$. Furthermore, $\vec z \sim \mathcal{N}(0, \Lambda)$, $\{E'_j\}$ satisfy the conditions for \cref{tm:vanishing}. \qed

Importantly \cref{tm:vanishing}, \cref{tm:vanishing-twopoint}, and \cref{tm:vanishing-nondiag} imply that, for tomographic protocols which sample errors as $\vec y\sim \mathcal{N}(0, \Sigma)$ where $\Sigma$ has eigenvalues $v_1,...,v_{D^2}$ for which $D\mathrm{var}[v]/\overline{v}$ vanishes, the integral of Lipschitz functions $f$ over the empirical measure $\mu_y$ ($\mu_y^{(2)}$) is identical at leading order to the integral of $f$ over a corresponding GUE measure $\mu_g$ ($\mu_y^{(2)}$) induced by $\vec g \sim \mathcal{N}(0, \overline{v}\mathbb{I})$ (via \cref{tm:pauli-theorem}). This powerful and flexible statement is the main result of our work.

We now show that $D\mathrm{var}[v]/\overline{v}$ vanishes at large system size $N$ and shot count $S$ for both a na\"ive toy tomographic protocol (see \cref{sec:naive}), and for a widely-used, more sophisticated tomographic protocol (see \cref{sec:sophisticated}). The first case corresponds to the simplest full tomography protocol and serves as a useful benchmark, whilst the latter has found widespread implementation in experiment (and is used commercially, e.g. in IBM's native state tomography functionality). In both of these cases, one can substitute the appropriate GUE ensemble for the object $\Delta\hat{\rho}$ in subsequent derivations of Lipschitz functions integrated over its spectrum.



\subsection{Na\"ive Independent Measurements Tomographic Protocol}
\label{sec:naive}

We here consider a na\"ive tomographic process in which each Pauli string $\hat P_j \in \mathcal{P}_N=\{\mathbb{I}, X, Y, Z\}^{\otimes N}$ is probed independently. For a $D=2^N$ dimensional system, each coefficient is determined by its respective measurement outcome list $X_{j}$ (see \cref{sec:likely}) which are each comprised of $S_j = S$ shots; such a na\"ive process requires that all $\sim SD^2=S4^N$ relevant `prepare-and-measure' circuits are evaluated.




\begin{theorem}\label{tm:naive-protocol}
    For the na\"ive tomographic protocol wherein $v_j = (1-\mathrm{Tr}(\hat{\rho} P_j)^2)/(SD^2)$:
    \begin{equation}\label{eq:naive-varcalc}
        D\frac{\mathrm{var}[v]}{\overline{v}} = \mathcal{O}\left(\frac{1}{SD^2}\right).
    \end{equation}
\end{theorem}

\textit{Proof}. The proof is a straightforward computational exercise: $\overline{v}$ is given by
\begin{equation}\label{eq:naive-mean}
\overline v = \frac{1}{D^2}\sum_j v_j = \frac{1}{SD^4}\left(D^2 - D\mathrm{Tr}(\rho^2)\right),
\end{equation}
such that the variance $\mathrm{var}[v] = (\sum_jv_j^2/D^2)-\overline{v}^2$ can be evaluated directly
\begin{equation}
    \mathrm{var}[v] = \frac{1}{S^2 D^6} \left(\sum_j\mathrm{Tr}(\hat{\rho}P_j)^4 - \mathrm{Tr}(\hat{\rho}^2)^2\right).
\end{equation}
We can now explicitly compute \cref{eq:naive-varcalc},
\begin{equation}\label{eq:naive-varcalc-2}
    D\frac{\mathrm{var}[v]}{\overline{v}} = \frac{1}{SD}\frac{\sum_j\mathrm{Tr}(\hat{\rho}P_j)^4 - \mathrm{Tr}(\hat{\rho}^2)^2}{D^2 - D\mathrm{Tr}(\hat{\rho}^2)}.
\end{equation}
Finally, we remark that $\sum_j \mathrm{Tr}(\hat{\rho}P_j)^4$ satisfies the inequality:
\begin{equation}
    \sum_j \mathrm{Tr}(\hat{\rho}P_j)^4 \leq \sum_j \mathrm{Tr}(\hat{\rho}P_j)^2 = D\mathrm{Tr}(\hat{\rho}^2) \leq D
\end{equation}
where we have used \cref{eq:theorem-trace-rule} and the fact $\mathrm{Tr}(\hat{\rho}^2) \leq 1$ for physical states $\hat{\rho}$. Substitution of the saturation value $D$ of the above inequality into \cref{eq:naive-varcalc-2} yields \cref{eq:naive-varcalc}; completing the proof. \qed

\subsection{QWC Tomographic Protocol}
\label{sec:sophisticated}

Whilst a convenient toy model of tomography, and a vital benchmark for our numerical and analytic results, the na\"ive protocol of \cref{sec:naive} is generally inefficient; one can obviously measure some Pauli strings simultaneously. Consider e.g. all Pauli strings comprised exclusively of local $Z$ operators and the identity, which can all be retrieved from the statistics of the individual local measurements used to construct the statistics for the all-$Z$ string $Z\otimes Z\otimes\cdots\otimes Z$. Thus more sophisticated QST protocols based on these combinations have been developed along these lines, taking advantage of what is referred to as qubit-wise commutativity (QWC) \cite{verteletskyi2020measurement, yen2020measuring}. For our Pauli measurement setup, QWC takes the following form. For every $Q\in\{X, Y, Z\}^{\otimes N}$, prepare a fixed number of $S$ copies $\rho$ and record measurement outcomes for every qubit. To construct the $4^N$ effective Pauli string measurements from these $3^N$ experiments: for a given Pauli string $P_j$ (i) collect every experiment $Q$ where $P_j|Q$; where $P_j|Q$ if, at every position where $P_j$ is non-identity, it matches with $Q$, (ii) discard the local measurement outcomes in $Q$ which corresponds to positions in $P_j$ which are the identity, (iii) agglomerate the remaining statistics and compute the coefficient for $P_j$ from \textit{all} experiments from which a measurement outcome for $P_j$ may be derived. This protocol generates a non-diagonal covariance matrix $\Sigma$ from which the Pauli coefficients $\vec{y}\sim\mathcal{N}(0,\Sigma)$ of the excess matrix $\Delta\hat\rho$ are sampled. We define $\delta^E_{ij} = 1$ if there is some $Q\in\{X, Y, Z\}^{\otimes N}$ such that $P_i|Q$ and $P_j|Q$, (i.e, $P_i$ and $P_j$ share data from some experiments), and $\delta^E_{ij} = 0$ otherwise. If $\delta^E_{ij} = 1$, $\Sigma_{ij}$ will in general be nonzero. The vector of coefficients is given by $\vec{y_{S}} \sim \mathcal{N}(0, \Sigma)$, with non-diagonal $\Sigma$.  To calculate $\Sigma_{ij}$ we must determine the number of experiments $Q$ that contributed to both $P_i$ and $P_j$. Defining the Pauli weight (number of non-identity single qubit entries in the string), of $P$ as $w(P)$; then $P$ is involved in $S3^{N-w(P)}$ experiments, and if $\delta^E_{ij} = 1$ then $P_i$ and $P_j$ share $S3^{w(P_i, P_j)}$ experiments, where $w(P, Q) = \max\{w(P), w(Q), w(PQ)\}$. Then,
\begin{equation}\label{eq:sophisticated-covariance}
    \Sigma_{ij} = k_{ij}\left[\mathrm{Tr}(\rho P_i P_j) - \mathrm{Tr}(\rho P_i)\mathrm{Tr}(\rho P_j)\right]/D^2,
\end{equation}
where the `trial ratio' $k_{ij}$ is given by:
\begin{equation}\label{eq:sophisticated-covariance-ratio}
    k_{ij} = \frac{\delta^E_{ij}}{S3^N}3^{w(P_i)+w(P_j) - w(P_i, P_j)}.
\end{equation}  
A comprehensive derivation of \cref{eq:sophisticated-covariance}, the trial ratio $k_{ij}$, and a discussion thereof, are given in \cref{sec:app-sigma-calc}.

\begin{theorem}\label{tm:sophisticated-protocol}
    For the QWC tomographic protocol with covariance matrix $\Sigma$ given by \cref{eq:sophisticated-covariance} (and thus correlated excess coefficients):
    \begin{equation}\label{eq:sophisticated-varcalc}
        D\frac{\mathrm{var}[v]}{\overline{v}} = \mathcal{O}\left(\frac{1}{S}\left(\frac{5}{192}\right)^N\right),
    \end{equation}
    where $v_j$ are the eigenvalues of $\Sigma$.
\end{theorem}

\textit{Proof.} To verify this, first note that $\overline v = \mathrm{Tr}(\Sigma)/D^2$. Thus:
\begin{widetext}
\begin{equation}\label{eq:sophisticated-mean0}
    \overline v = \frac{1}{SD^4}\sum_j\frac{1-\mathrm{Tr}(\rho P_j)^2}{3^{N-w(P_j)}} = \frac{1}{SD^43^N}\left(\sum_j3^{w(P_j)} - \sum_j 3^{w(P_j)}\mathrm{Tr}(\rho P_j)^2\right).
\end{equation}
\end{widetext}
To evaluate the first term on the r.h.s., note that there are ${10\choose k}3^k$ Pauli strings with weight $k$, and hence:
\begin{equation}\label{eq:sophisticated-mean0-a} 
    \sum_j3^{w(P_j)} = \sum_{k=0}^N {10\choose k}9^k = 10^N.
\end{equation}
The second term on the r.h.s. vanishes at leading order in $N$:
\begin{equation}\label{eq:sophisticated-mean0-b} 
    \sum_j 3^{w(P_j)}\mathrm{Tr}(\rho P_j)^2 \leq 3^N \sum_j\mathrm{Tr}(\rho P_j^2) \leq 6^N.
\end{equation}
Direct substitution of \cref{eq:sophisticated-mean0-a} and \cref{eq:sophisticated-mean0-b} into \cref{eq:sophisticated-mean0} yields,
\begin{equation}\label{eq:sophisticated-mean}
 \overline{v} = \frac{1}{S}\left(\frac{5}{24}\right)^N +\mathcal{O}\left(\left(\frac{3}{24}\right)^N\right).
\end{equation}
We now turn our attention to $\mathrm{var}[v]$:
\begin{widetext}
\begin{equation}
    \mathrm{var}[v] = \frac{1}{D^2}\mathrm{Tr}(\Sigma^2) - \left(\frac{1}{D^2}\mathrm{Tr}(\Sigma)\right)^2 \leq \frac{1}{D^2}\mathrm{Tr}(\Sigma^2) = \frac{1}{D^2}\|\Sigma\|_F^2.
\end{equation}
\end{widetext}
Let $\Gamma_{ij} =(\mathrm{Tr}(\rho P_i P_j) - \mathrm{Tr}(\rho P_i)\mathrm{Tr}(\rho P_j))/D^2$. Then,
\begin{equation}
    \|\Sigma\|_F^2 = \sum_{ij}|k_{ij}|^2|\Gamma_{ij}|^2 \leq(\mathrm{max}[k_{ij}])^2\|\Gamma\|_F^2
\end{equation} 
Remarking that $\|\Gamma\|_F^2\leq 4/D^2$, and $\max[k_{ij}]\leq 1/S$, then $\|\Sigma\|_F^2 \leq 4/(SD)^2$ and thus
\begin{equation}\label{eq:sophisticated-var}
    \mathrm{var}[v]\leq \frac{4}{S^2 D^4}.
\end{equation}
Direct use of \cref{eq:sophisticated-mean} and \cref{eq:sophisticated-var} yields:

\begin{equation}
    \frac{D\mathrm{var}[v]}{\overline v} = \mathcal{O}\left(\frac{1}{S}\left(\frac{5}{192}\right)^N\right),
\end{equation}

and thus \cref{tm:vanishing-nondiag} implies that samples drawn from the GUE measure can be substituted for realizations of $\mathcal{A}_P(\vec{z})$ in the calculation of Lipschitz functions over the one-point and two-point empirical spectral measures, with small corrections at large $N, S$.

\section{Results}
\label{sec:results}

Here we present results on a variety of topics in QST which are direct consequences of the preceding analyses. Firstly, in \cref{sec:errors} we derive rigorous, yet simple, analytic forms for the expected values $\mathbb{E}[\|\Delta\hat\rho\|_\mathrm{Tr}]$ and variances $\mathrm{var}[\|\Delta\hat\rho\|_\mathrm{Tr}]$ of the QST error (excess trace distance) $\|\Delta\hat\rho\|_\mathrm{Tr}$ for any tomographic protocols satisfying \cref{tm:vanishing} (and thus \cref{tm:vanishing-twopoint}) and with vanishing $D\mathrm{var}[v]/\overline{v}$, to leading order in $S$ and $D$. In \cref{sec:numerics} we numerically validate these rigorous forms for both the na\"ive and QWC tomographic protocols discussed in \cref{sec:naive} and \cref{sec:sophisticated} respectively. In \cref{sec:complexity} we validate a recently derived bound on the sample complexity using only RMT methods; supporting the conjecture of the authors of Ref.~\cite{acharya2025pauli} that said bound is tight, at least for non-adaptive QST. In \cref{sec:rephysicalization} we demonstrate that the most widely used rephysicalization algorithm, a maximum-likelihood projection of $\hat{\rho}^\prime$ back to the physical subspace, does not strengthen the bound of \cref{eq:bound}; i.e. one cannot use an additional physical constraint of this form to improve the sample complexity of the problem $\|\hat{\rho}-\hat{\rho}^\prime\|_\mathrm{Tr}\leq\varepsilon$. Finally, in \cref{sec:speculate}, we briefly discuss several speculative avenues of future research.

\subsection{Analytic Results for Quantum State Tomography Errors}
\label{sec:errors}

We turn our attention to the central object of inquiry of this work, namely the trace distance
\begin{equation}\label{eq:trdist}
    \|\hat{\rho} - \hat{\rho}^\prime\|_\mathrm{Tr} = \|\Delta\hat\rho\|_\mathrm{Tr} = \sum_{j}|\lambda_j(\Delta\hat\rho)|,
\end{equation}
between the ontological state $\hat{\rho} \in \mathrm{phys}_D$ and its (potentially aphysical) tomographic reconstruction $\hat{\rho}^\prime \in \mathrm{hut}_D$. We here calculate expected values $\mathbb{E}[\|\Delta\hat\rho\|_\mathrm{Tr}]$ and variances $\mathrm{var}[\|\Delta\hat\rho\|_\mathrm{Tr}]$ of this trace distance for both the na\"ive and QWC tomographic protocols discussed in \cref{sec:naive} and \cref{sec:sophisticated} respectively. In both of these cases we have shown that $D\mathrm{var}[v]/\overline{v}$ vanishes with $D$ and $S$ (see \cref{tm:naive-protocol} and \cref{tm:sophisticated-protocol} respectively), and thus that \cref{tm:vanishing}, \cref{tm:vanishing-twopoint}, and \cref{tm:vanishing-nondiag} imply that at leading order, samples from the GUE ensemble can be substituted for samples of $\Delta\hat{\rho}$ for calculating Lipschitz functions over one-point and two-point empirical spectral measures. We note here that $\mathbb{E}[\|\Delta\hat\rho\|_\mathrm{Tr}]$ and $\mathrm{var}[\|\Delta\hat\rho\|_\mathrm{Tr}]$ are two such functions; let $f(x) = |x|$ (Lipschitz), and let $\Delta\hat\rho = \mathcal{A}_P(\vec y)$ with empirical one-point and two-point measures given by \cref{eq:one-point} and \cref{eq:two-point} respectively, then the mean $\mathbb{E}[\|\Delta\hat\rho\|_\mathrm{Tr}]$ is given by:
\begin{equation}
    \mathbb{E}[\|\Delta\hat\rho\|_\mathrm{Tr}] = \mathbb{E}\left[\int f \,\mathrm{ d}\mu_y\right],
\end{equation}
and the variance $\mathrm{var}[\|\Delta\hat\rho\|_\mathrm{Tr}]$ by 
\begin{widetext}
\begin{equation}
    \mathrm{var}[\|\Delta\hat\rho\|_\mathrm{Tr}] = \mathbb{E}\left[\int f\otimes f \,\mathrm{d}\mu_y^{(2)}\right] +\mathbb{E}\left[\int f^2\,\mathrm{d}\mu_y \right] -\mathbb{E}\left[\int f\,\mathrm{d}\mu_y\right]^2,
\end{equation}
\end{widetext}
which both consist of (expected values of) Lipschitz functions of the one-point and two-point measures. Thus, by \cref{tm:vanishing} and \cref{tm:vanishing-twopoint}, for vanishing $D\mathrm{var}[v]/\overline{v}$ as shown for the na\"ive and QWC tomographic protocols in \cref{tm:naive-protocol} and \cref{tm:sophisticated-protocol} respectively, we can substitute the measure induced by $\mathcal{A}_E(\vec{y})$ with the GUE measure to leading order in $S$ and $N$. This in turn allows us to evaluate $\mathbb{E}[\|\Delta\hat\rho\|_\mathrm{Tr}]$ and $\mathrm{var}[\|\Delta\hat\rho\|_\mathrm{Tr}]$ (to leading order) exactly.

\begin{theorem}
    For tomographic protocols satisfying \cref{tm:vanishing}, and with vanishing $D\mathrm{var}[v]/\overline{v}$; to leading order in $D$ and $S$ the expected value of the trace distance between the ontological state $\hat{\rho}$ and it's tomographically determined counterpart $\hat{\rho}^\prime$ is given by
    \begin{equation}\label{eq:mean}
        \mathbb{E}[\|\Delta\hat{\rho}\|_\mathrm{Tr}] = \frac{4D}{3\pi}R
    \end{equation}
    where $R = 2D\sqrt{\overline{v}}$ is the Wigner radius for measures of the form of \cref{eq:pauli-measure}, and where $\overline{v}$ is determined by the tomographic protocol.
\end{theorem}
\textit{Proof.} The proof proceeds by direct evaluation of $\mathbb{E}[\|\Delta\hat\rho\|_\mathrm{Tr}]$ over the Wigner semi-circle distribution:
\begin{equation}
    \mathbb{E}[\|\Delta\hat\rho\|_\mathrm{Tr}] = D\int_{-R_S}^{R_S}|\lambda|W_R(\lambda)d\lambda.
\end{equation}
Which can be evaluated by exploiting the fact that $W_R(\lambda)$ is an even function:
\begin{equation}
    \mathbb{E}[\|\Delta\hat\rho\|_\mathrm{Tr}] = 2D\int_{0}^{R_S}\lambda W_R(\lambda)d\lambda= \frac{4D}{3\pi}R,
\end{equation}
completing the proof. \qed 
\begin{theorem}
    For tomographic protocols satisfying \cref{tm:vanishing} (and thus \cref{tm:vanishing-twopoint}) and with vanishing $D\mathrm{var}[v]/\overline{v}$; to leading order in $D$ and $S$ the variance $\mathrm{var}[\|\Delta\hat{\rho}\|_\mathrm{Tr}]$ of the distance $\Delta\hat{\rho} = \hat{\rho}-\hat{\rho}^\prime$ between the ontological state $\hat{\rho}$ and it's tomographically determined counterpart $\hat{\rho}^\prime$ is given by
    \begin{equation}\label{eq:var}
        \mathrm{var}[\|\Delta\hat{\rho}\|_\mathrm{Tr}] = \frac{R^2}{\pi^2} 
    \end{equation}
    where $R = 2D\sqrt{\overline{v}}$ is the Wigner radius for measures of the form of \cref{eq:pauli-measure}, and where $\overline{v}$ is determined by the tomographic protocol.
\end{theorem}
\textit{Proof.} Using the results of Ref.~\cite{chen1998linear}, notably Eq.~(2.22) therein, the variance $\mathrm{var}[\|\Delta\hat{\rho}\|_\mathrm{Tr}]$ can be calculated as
\begin{widetext}
\begin{equation}\label{eq:full-variance}
    \mathrm{var}[\|\Delta\hat{\rho}\|_\mathrm{Tr}] = \frac{1}{2\pi^2}\int_{-R_S}^{R_S}\mathrm{d}\lambda_1\frac{|\lambda_1|}{\sqrt{R_S^2-\lambda_1^2}}\mathcal{P}\left[\int_{-R_S}^{R_S}\mathrm{d}\lambda_2\frac{\sqrt{R_S^2-\lambda_2^2}}{\lambda_1-\lambda_2}\mathrm{sgn}(\lambda_2)\right] = \frac{\mathcal{I}_+ + \mathcal{I}_-}{\pi^2}
\end{equation}
\end{widetext}
where $\mathcal{P}[\int^b_a\cdot]$ denotes the principal value of the (improper) integral. The objects $\mathcal{I}_\pm$ are retrieved by explicitly considering two different regimes: $\mathrm{sgn}(\lambda_1)=\mathrm{sgn}(\lambda_2)$ and $\mathrm{sgn}(\lambda_1)=-\mathrm{sgn}(\lambda_2)$, yielding:
\begin{equation}\label{eq:improper-terms}
    \mathcal{I}_{\pm} = \pm \int^R_0\mathrm{d}\lambda_1\mathcal{P}\left[\int^R_0\mathrm{d}\lambda_2\frac{\lambda_1}{\lambda_1\mp\lambda_2} \sqrt{\frac{R^2-\lambda_2^2}{R^2-\lambda_1^2}}\right]
\end{equation}
respectively. We can explicitly compute $\mathcal{I}_\pm = R^2(4\pm \pi^2)/8$ by factorizing the integrand into a form to which generic integral formulae apply, and taking limits appropriately to resolve the principal value. Direct substitution into \cref{eq:full-variance} yields the variance
\begin{equation}
    \mathrm{var}[\|\Delta\hat\rho\|_\mathrm{Tr}] = \frac{R^2}{\pi^2}
\end{equation}
for Wigner radius $R = 2D\sqrt{\overline{v}}$, completing the proof. \qed


\begin{figure*}
    \centering
    \includegraphics[width=1.0\linewidth]{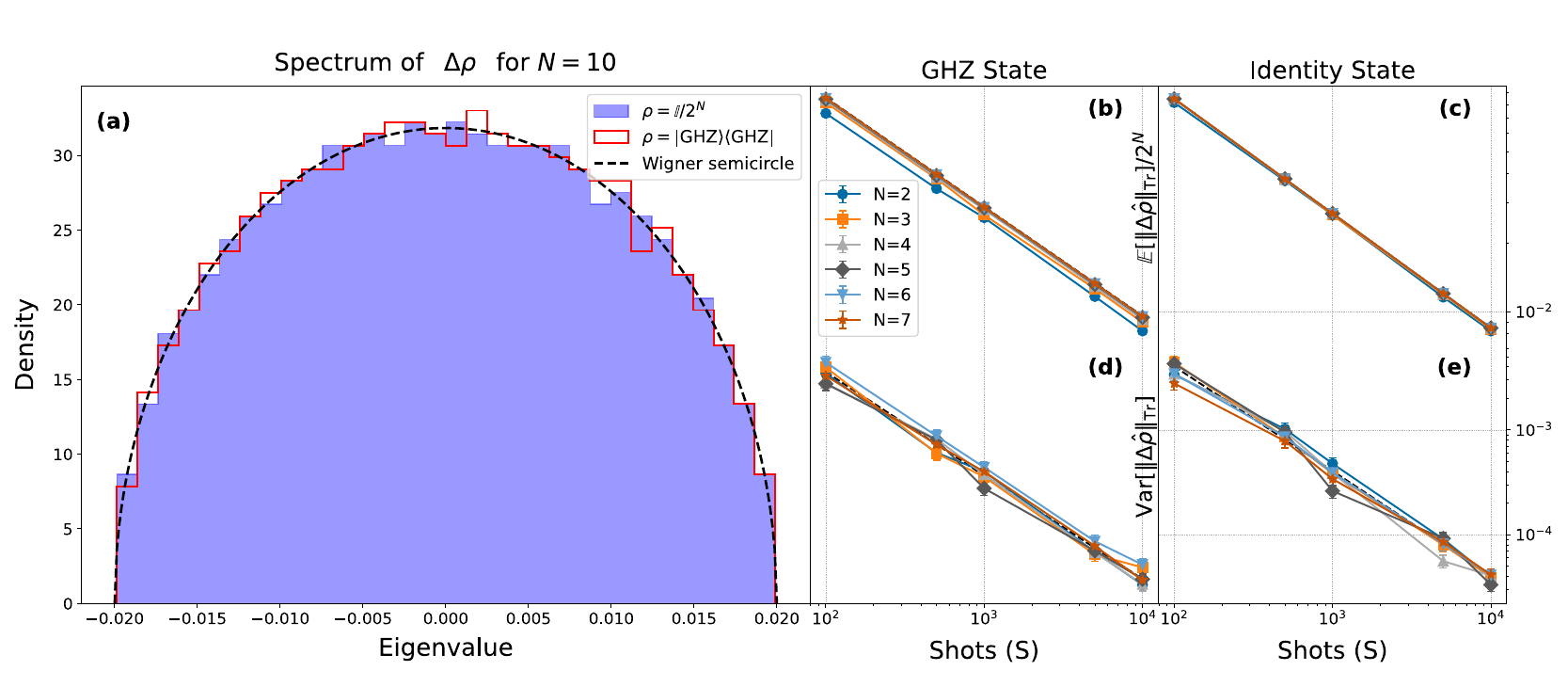}
    \caption{Numerical analyses of the na\"ive procedure applied to the GHZ and identity state. System size $N=10$ and $S=10^4$ shots taken throughout. \textbf{(a)} shows the eigenvalue distribution of a single realisation of $\Delta\hat\rho$ with the Wigner semicircle of radius $R=2/\sqrt{S}$ superimposed. \textbf{(b)} and \textbf{(c)} show numerically determined (rescaled) expectation values $\mathbb{E}[\|\Delta\hat\rho\|_\mathrm{Tr}] / 2^N$ with the derived analytic prediction (\cref{eq:naive-res-mean}) superimposed for the GHZ and identity state respectively; \textbf{(d)} and \textbf{(e)} show the corresponding variances $\mathrm{var}[\|\Delta\hat\rho\|_\mathrm{Tr}] 2^N$, again with the analytic prediction (\cref{eq:naive-res-var}) superimposed. Panels \textbf{(b)}-\textbf{(e)} evaluated on 100 realizations of $\Delta \hat \rho$, with error bars shown where visible.}
    
    \label{fig:naive-results}
\end{figure*}

\subsection{Independent and Correlated Coefficients}
\label{sec:numerics}

\begin{figure*}
    \centering
    \includegraphics[width=\linewidth]{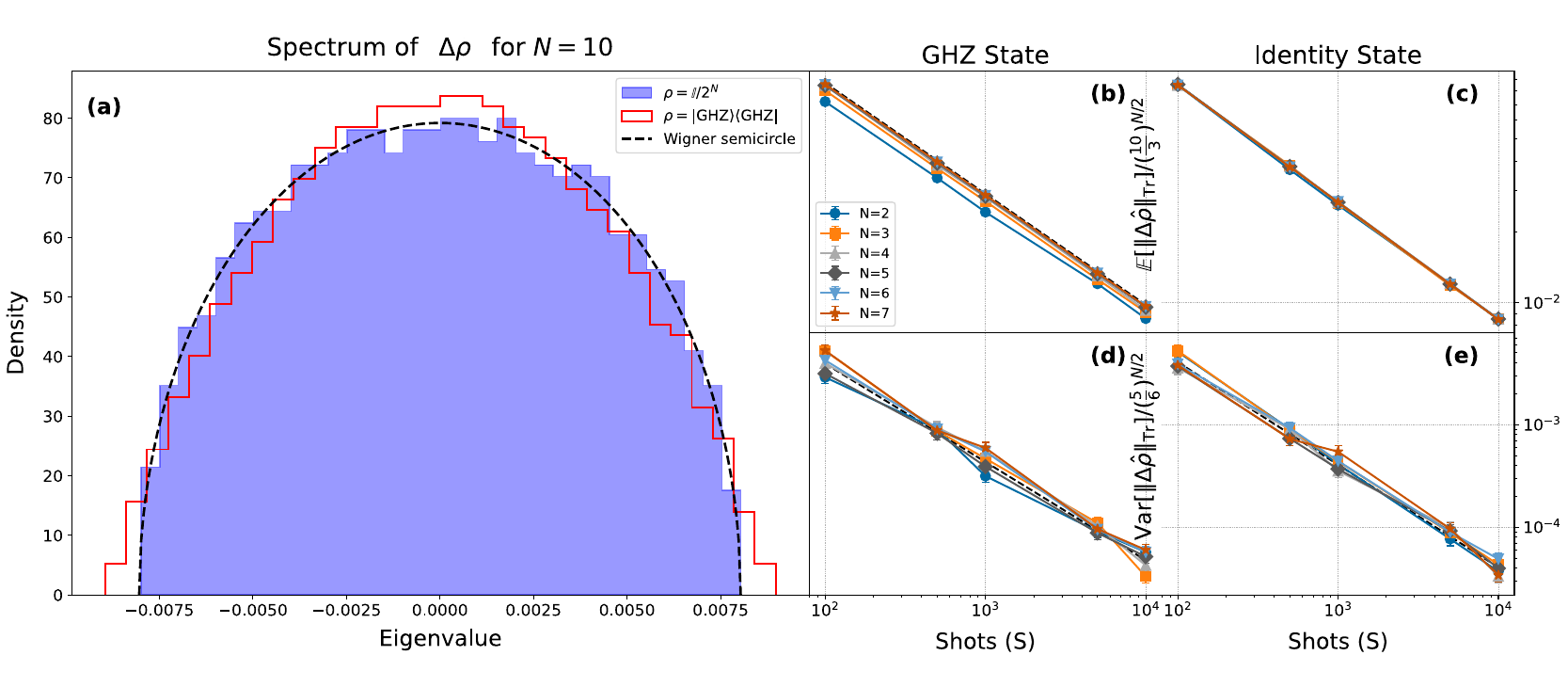}
    \caption{Numerical analyses of the QWC procedure applied to the GHZ and identity state. System size $N=10$ and $S=10^4$ shots taken throughout. \textbf{(a)} shows the eigenvalue distribution of a single realisation of $\Delta\hat\rho$ with the Wigner semicircle of radius $R=2/\sqrt{S}$ superimposed; with the GHZ state showing some deformation away from GUE level statistics. \textbf{(b)} and \textbf{(c)} show numerically determined (rescaled) expectation values $\mathbb{E}[\|\Delta\hat\rho\|_\mathrm{Tr}] / (10/3)^{N/2}$ with the derived analytic prediction (\cref{eq:soph-res-mean}) superimposed for the GHZ and identity state respectively; \textbf{(d)} and \textbf{(e)} show the corresponding variances $\mathrm{var}[\|\Delta\hat\rho\|_\mathrm{Tr}] 2^N$, again with the analytic prediction (\cref{eq:soph-res-var}) superimposed. We remark on the excellent agreement of the GHZ state numerics with GUE-derived analytic predictions in particular, despite the disagreement between GHZ-sampled and Wigner semicircle level statistics shown in panel \textbf{(a)}; this is a concrete realization of our \cref{tm:vanishing}. Panels \textbf{(b)}-\textbf{(e)} evaluated on 100 realizations of $\Delta \hat \rho$, with error bars shown where visible.}
    \label{fig:soph-results}
\end{figure*}

In the na\"ive case of independent experiments (see \cref{sec:naive}) $\overline{v}$ is given by \cref{eq:naive-mean} which, to leading order in $S$ and $D$, gives $\overline{v} = 1/SD^2$ and a corresponding effective Wigner radius of $R = 2/\sqrt{S}$ (via \cref{eq:wigner-radius}). This radius yields, using \cref{eq:mean} and \cref{eq:var} respectively, mean value and variance
\begin{align} \label{eq:naive-res-mean}
    \mathbb{E}[\|\Delta\hat\rho\|_\mathrm{Tr}] &= \frac{8D}{3\pi\sqrt{S}}, \\ \qquad\mathrm{var}[\|\Delta\hat\rho\|_\mathrm{Tr}] &= \frac{4}{\pi^2 S}.\label{eq:naive-res-var}
\end{align}
In the case of correlated coefficients for the QWC tomographic protocol, $\overline{v}$ is given by \cref{eq:sophisticated-mean} which, to leading order in $S$ and $N$, gives $\overline{v} = (5/24)^N/S$ and a corresponding effective Wigner radius of $R = 2(5/6)^{N/2}/\sqrt{S}$ (via \cref{eq:wigner-radius}). This radius yields, using \cref{eq:mean} and \cref{eq:var} respectively, mean value and variance
\begin{align}\label{eq:soph-res-mean}
    \mathbb{E}[\|\Delta\hat\rho\|_\mathrm{Tr}] &= \frac{8}{3\pi\sqrt{S}}\left(\frac{10}{3}\right)^{N/2}, \\ \qquad\mathrm{var}[\|\Delta\hat\rho\|_\mathrm{Tr}] &= \frac{4}{\pi^2 S} \left(\frac{5}{6}\right)^N.\label{eq:soph-res-var}
\end{align}
We validate these mean values and variances against numerical data in \cref{sec:numerics} in \cref{fig:naive-results} and \cref{fig:soph-results} respectively, which we discuss here.

First consider the na\"ive tomographic protocol of \cref{sec:naive} for realizátion(s) of $\Delta\hat\rho$ under said protocol using a shot count of $S=10^4$, sampling the GHZ state $|\mathrm{GHZ}\rangle\langle\mathrm{GHZ}|$ where $|\mathrm{GHZ}\rangle = (|0\rangle^{\otimes N}+|1\rangle^{\otimes N})/\sqrt{2}$, and identity $\mathbb{I}/2^N$ respectively; where we have taken $N=10$ throughout. We have selected these states to highlight certain aspects of our RMT approach, and (in the case of the highly correlated and structured GHZ state) to eliminate any possibility that internal state averaging leads to random matrix statistics.

In \cref{fig:naive-results}\textbf{(a)} we show the eigenvalue distributions of a single realization of $\Delta\hat\rho$ for both states respectively, and have superimposed the Wigner semicircle distribution of radius $R=2/\sqrt{S}$ (using the na\"ive protocol's value for $\overline{v}$) on \cref{fig:naive-results}\textbf{(a)} ; and can clearly see agreement between all three sets of data. This suggests an even stronger statement than those made in \cref{sec:likely}; wherein for the case of na\"ive QST in the Pauli IC-basis, the ensemble of $\Delta\hat\rho$ itself converges to the GUE. Panels \textbf{(b)} and \textbf{(c)} show the numerically determined rescaled expectation values $\mathbb{E}[\|\Delta\hat\rho\|_\mathrm{Tr}] / 2^N$ for the GHZ and identity states respectively, and panels \textbf{(d)} and \textbf{(e)} show the corresponding variances $\mathrm{var}[\|\Delta\hat\rho\|_\mathrm{Tr}]$. We have taken 100 sample realizations of $\Delta\hat\rho$ for every data point, and have superimposed the (rescaled) analytic values of \cref{eq:naive-res-mean} and \cref{eq:naive-res-var}. We find strong agreement at all system sizes; with slight deviation at $N=2$ for the GHZ state due to the fact that our main results (and thus derived expected values and variances) hold at leading order in $N$. The reason that the identity state does not exhibit this system-size dependence, and the reason that it shows almost exact agreement with our analytic prediction, is because all the Pauli coefficients (excluding the trivial all-identity Pauli string) are zero, and thus induce an almost-uniform diagonal $\Sigma$ - allowing us to apply \cref{eq:pauli-measure} directly.


Second consider the sophisticated tomographic protocol of \cref{sec:sophisticated} for realizátion(s) of $\Delta\hat\rho$ under said protocol, sampling the GHZ and identity states respectively. Again we take $N=10$ and $S=10^4$ throughout.

In \cref{fig:soph-results}\textbf{(a)} we show the eigenvalue distributions of a single realization of $\Delta\hat\rho$ for both states respectively, and have superimposed the Wigner semicircle distribution of radius $R=2(5/6)^{N/2}/\sqrt{S}$ (using the sophisticated protocol's value for $\overline{v}$) on \cref{fig:soph-results}\textbf{(a)}. In contrast to the results for the na\"ive protocol, the eigenvalue distribution for the GHZ state does not uniformly converge to the Wigner semicircle distribution derived from the GUE measure. This is due to the covariances (off-diagonal elements in $\Sigma$, see \cref{eq:sophisticated-covariance}) that arise due to cross-qubit correlations. Despite this, an in accordance with our \cref{tm:vanishing}, \cref{tm:vanishing-nondiag}, and \cref{tm:vanishing-twopoint}, the behaviour of functions of one-point and two-point empirical spectral measures of the error in GHZ state sampling should still agree with GUE predictions; and indeed this is precisely the behaviour shown in panels \textbf{(b)} and \textbf{(d)}. Note sampling the identity state using the QWC protocol still leads to diagonal $\Sigma$ by \cref{eq:sophisticated-covariance} as there are no cross-qubit correlations in the a priori state. Panels \textbf{(b)} and \textbf{(c)} show the numerically determined rescaled expectation values $\mathbb{E}[\|\Delta\hat\rho\|_\mathrm{Tr}] / (10/3)^{N/2}$ for the GHZ and identity states respectively, and panels \textbf{(d)} and \textbf{(e)} show the corresponding rescaled variances $\mathrm{var}[\|\Delta\hat\rho\|_\mathrm{Tr}] / (5/6)^{N/2}$. Again we have taken 100 sample realizations of $\Delta\hat\rho$ for every data point, and have superimposed the (rescaled) analytic values of \cref{eq:soph-res-mean} and \cref{eq:soph-res-var}. Once more we find excellent agreement at all system sizes with slight deviation $N=2$ for the GHZ state. This provides direct numerical support for our main result of \cref{sec:likely} that GUE samples can be substituted for samples of $\Delta\hat\rho$ in the calculation of functions of the one-point and two-point empirical spectral measures; even when said spectra pointwise disagree with respective GUE predictions.

\subsection{Non-Adaptive Pauli Tomography Sample Complexity}
\label{sec:complexity}

We here derive the sample complexity for the QWC non-adaptive Pauli tomographic scheme outlined in \cref{sec:sophisticated}; in doing so we want to bound the probability that expected distance $\|\Delta\hat\rho\|_\mathrm{Tr} = \|\hat\rho - \hat\rho^\prime\|_
\mathrm{Tr}$ between the ontological state $\hat{\rho}$ and the tomographically determined reconstruction $\hat{\rho}^\prime$ exceeds some threshold parameter $\varepsilon$, which is precisely given by Markov's inequality:
\begin{equation}
    \mathbb{P}\left[\|\Delta\hat\rho\|_\mathrm{Tr} \geq \varepsilon\right] \leq \frac{8}{3\pi\varepsilon\sqrt{S}}\left(\frac{10}{3}\right)^{N/2}
\end{equation}
where we have used the value of $\mathbb{E}[\|\Delta\hat\rho\|_\mathrm{Tr}]$ given in \cref{eq:sophisticated-mean}. To guarantee that this distance is not exceeded with probability $p$, a shot count of
\begin{equation}
    S \geq \left(\frac{8}{3\pi\varepsilon p}\right)^2\left(\frac{10}{3}\right)^N
\end{equation}
is sufficient for each of the $3^N$ distinct measurement setups. This yields a sample complexity of
\begin{equation}
    \label{eq:bound}
    n \sim \mathcal{O}\left(\frac{10^N}{\varepsilon^2}\right),
\end{equation}
where $n$ is the total number of copies of $\hat\rho$ that must be prepared for the tomographic reconstruction $\hat\rho^\prime$ to have a fixed probability $p$ of being within a trace distance $\varepsilon$ of the ontological state $\hat\rho$.

\begin{theorem}
    The bound for the sample complexity given in \cref{eq:bound} is tight.
\end{theorem}
\textit{Proof.} We can make this a tight bound if we consider the Paley–Zygmund inequality. For $0<\varepsilon<\mathbb{E}[\|\Delta\hat\rho\|_\mathrm{Tr}]$ we have:
\begin{equation}
    \mathbb{P}[\|\Delta\hat\rho\|_\mathrm{Tr} \geq \varepsilon] \geq  \frac{\left(\mathbb{E}[\|\Delta\hat\rho\|_\mathrm{Tr}]-\varepsilon\right)^2}{\mathbb{E}[\|\Delta\hat\rho\|_\mathrm{Tr}]^2+\mathrm{var}[\|\Delta\hat\rho\|_\mathrm{Tr}]}.
\end{equation}
First, note that $\mathrm{var}[\|\Delta\hat\rho\|_\mathrm{Tr}]/\mathbb{E}[\|\Delta\hat\rho\|_\mathrm{Tr}]^2\longrightarrow 0$ as $N\rightarrow \infty$, and so for large $N$:
\begin{equation}\label{eq:pz2}
    \mathbb{P}[\|\Delta\hat\rho\|_\mathrm{Tr} \geq \varepsilon]\geq \left(1-\frac{\varepsilon}{\mathbb{E}[\|\Delta\hat\rho\|_\mathrm{Tr}]}\right)^2 (1+ o(1))
\end{equation}
Now to show that $S\sim \left(\frac{10}{3}\right)^N/\varepsilon^2$ is not only sufficient, but necessary to ensure any upper bound $p$ on the failure probability ($\mathbb{P}[\|\Delta\hat\rho\|_\mathrm{Tr}\geq \varepsilon] \leq p$), let's suppose otherwise. Suppose that for any $c>0$, we have $S\leq c\left(\frac{10}{3}\right)^N/\varepsilon^2$ for large enough $N$. Let the precision $\varepsilon$ be fixed and choose $c\leq(4/3\pi)^2$. Then for large enough $N$, using \cref{eq:soph-res-mean} and our above assumption:
\begin{equation}
    \frac{\mathbb{E}[\|\Delta\hat\rho\|_\mathrm{Tr}]}{\varepsilon} \geq 2.
\end{equation}
Direct substitution into \cref{eq:pz2} yields
\begin{equation}
    \mathbb{P}[\|\Delta\hat\rho\|_\mathrm{Tr} \geq \varepsilon] \geq \left(1-\frac{1}{2}\right)^2 (1+o(1)),
\end{equation}
and so the failure probability has a floor for large $N$: \begin{equation}
    \mathbb{P}[\|\Delta\hat\rho\|_\mathrm{Tr} \geq \varepsilon]\gtrsim 1/4.
\end{equation}
Thus, our assumption that it is sufficient for $S$ to grow slower than $(10/3)^N/\varepsilon^2$ to ensure any upper bound $p$ on the failure probability is incorrect. Using this with \cref{eq:bound}, we get:
\begin{equation}
    n\sim \Theta\left(\frac{10^N}{\varepsilon^2}\right).
\end{equation}
\qed

We have shown that our unbiased estimator for the QWC tomographic setup confirms the upper bounding of the sample complexity for single-copy Pauli QST in \cite{acharya2025pauli} and moreover, proves the tightness of the bound in this setup. This recently found complexity bound pops up solely with a RMT treatment of the Pauli error matrix $\Delta\hat\rho$, with no heavy machinery from quantum information theory. In \cref{sec:rephysicalization} we discuss the possible generalization of this result to any estimator based on non-adaptive Pauli QST. 

\subsection{Comparison to the Maximum Likelihood Estimator}
\label{sec:rephysicalization}

Consider the estimator constructed in Ref.~\cite{smolin2012efficient}, constructed by rephysicalizing the estimator $\hat{\rho}^\prime$, by setting the negative eigenvalues to zero and `shifting' the non-zero elements of the spectrum in such a way to enforce the unit trace condition. This can be understood as a maximum-likelihood (i.e. distance-minimizing) projection of $\hat{\rho}^\prime \in\mathrm{hut}_D$ back onto the space $\mathrm{phys}_D$ of physical states (as shown schematically in \cref{fig:schematic}). Let $\hat\Pi \hat\rho^\prime \in \mathrm{phys}_D$ denote the resulting projected state, and let $\mu$ and $\mu'$ be the empirical spectral measures of the ontological state $\hat\rho$ and reconstructed state $\hat\rho^\prime$ respectively. Then the corresponding eigenvalue probability density functions are given by $\mathrm{d}\mu/\mathrm{d}\lambda, \mathrm{d}\mu^{\prime}/\mathrm{d}\lambda$. It follows then from the fact that the distance $\|\hat\Pi \hat\rho^\prime - \hat\rho^\prime\|_\mathrm{Tr}$ is minimized by enforcing mutual diagonalizability of $\hat\rho^\prime$ and $\hat\Pi \hat\rho^\prime$, and from the procedure of Ref.~\cite{smolin2012efficient}, that:
\begin{equation}
\label{eq:rephys}
    \|\hat\Pi \hat\rho^\prime - \hat\rho^\prime\|_\mathrm{Tr} = 2\left|\int_{-R_S}^0 |\lambda| \frac{d\mu'}{d\lambda}d\lambda\right|.
\end{equation}
Thanks to \cref{tm:vanishing-nondiag} and \cref{tm:sophisticated-protocol}, for the purpose of evaluating \cref{eq:rephys} to leading order in $S$ and $D$, we can treat the eigenvalue probability density function as the convolution of the underlying ontological density function with the Wigner semicircle:
\begin{equation}\label{eq:convolution}
    \frac{d\mu'}{d\lambda}(\lambda) = \left[\frac{d\mu}{d\lambda}\ast W_{R}\right ](\lambda),
\end{equation}
where $\ast$ denotes convolution, and $W_R$ is defined in \ref{eq:wigner} for Wigner radius $R$. We can now rewrite \cref{eq:rephys} as
\begin{equation}
\label{eq:smear}
   \|\hat\Pi \hat\rho^\prime - \hat\rho^\prime\|_\mathrm{Tr} =\left| \int^{R}_0 \frac{d\mu}{d\lambda} f(\lambda)\,d\lambda\right|,
\end{equation}
where $f(\lambda)$ extracts the value of the integral where the Wigner semi-circle has support on the negative eigenvalues, and is given by:
\begin{equation}
   f(\lambda) = 2\int^0_{\lambda-R}\nu W_S(\nu-\lambda)\,d\nu.
\end{equation}
As $\mu$ only has support on the non-negative eigenvalues, contributions to \cref{eq:smear} are only possible occur where $W_R(\lambda)$ is negative. \cref{eq:smear} is largest when the spectrum of $\hat\rho$ is peaked at the argmax of $|f(\mu)|$, and gives the bound
\begin{equation}
    \|\hat\Pi \hat\rho^\prime - \hat\rho^\prime\|_\mathrm{Tr} \leq 2R = \frac{2}{\sqrt{S}}\left(\frac{5}{6}\right)^{N/2}.
\end{equation}
Therefore we have that 
\begin{equation}
    \|\hat\Pi \hat\rho^\prime - \hat\rho\|_\mathrm{Tr} \geq \left|\|\Delta\hat\rho\|_\mathrm{Tr} - \|\hat\Pi \hat\rho^\prime - \hat\rho\|_\mathrm{Tr} \right|.
\end{equation}
The first term in the absolute value sign on the left is of order $\left(10/3\right)^{N/2}/\sqrt{S}$, while the second is bounded by a term of order $\left(5/6\right)^{N/2}/\sqrt{S}$. From this lower bound, and the fact that $\hat\Pi$ is contractive, we must have:
\begin{equation}\label{eq:complexity}
    \mathbb{E}[\|\hat\Pi \hat\rho^\prime- \hat\rho\|_\mathrm{Tr}] \sim \Theta\left(\frac{1}{\sqrt{S}}\left(\frac{10}{3}\right)^{N/2}\right),
\end{equation}
i.e, the maximum likelihood protocol does not improve the sample complexity of the bare QWC tomographic protocol as discussed in \cref{sec:complexity}. 

We believe but are unable to show in this work that \textit{any} rephysicalization procedure necessarily cannot improve the sample procedure of tomographic reconstruction, since it will involve mapping an object from $\mathrm{hut}_D$ to a convex subset of linear codimension $0$. We offer the above result in support of this claim, with the view that \cref{eq:complexity} holds for any single-copy, non-adaptive QST estimator for the a priori $\hat\rho$.

\subsection{Other Directions}
\label{sec:speculate}

As \cref{eq:bound} matches up with a recently found upper bound for Pauli tomography in \cite{acharya2025pauli}, it would be safe to guess that random matrix tools have potential to be useful in studying and assisting tomography. In particular, the generalisation of such analyses to \cref{eq:ic} and \cref{eq:sicpovm} would allow a wider range of tomographic protocols to be analysed as we have done for the Pauli case, and although we assumed qubit systems from the start for ease of exposition, \cref{eq:ic} clearly holds in for POVMs in general.

We have demonstrated in \cref{tm:vanishing}, \cref{tm:vanishing-twopoint}, \cref{tm:vanishing-nondiag} that the discussed tomographic protocols yield GUE results at leading order for Lipschitz functions over the one and two point empirical spectral measures (and as an easy generalisation, any $k$-point empirical spectral measure where $k$ is fixed for growing $N$); whether this correspondence holds for other functions of the spectrum of $\Delta\hat{\rho}$ - and what functions would be useful - may be fruitful questions.

Gaussian states by definition have a clear advantage in this framework, and it would be interesting to use GUE methods to study tomographic setups for homodyne detection. In particular, how finite sample reconstructions of the covariance matrix of a Gaussian state are expected to deviate from the actual covariance matrix, and what this can say about trace distance errors in Gaussian state reconstructions \cite{holevo2024estimates, zhang2023trace}. 

\section{Conclusions}
\label{sec:conclusions}

In this work, we have developed a fully random matrix treatment of QST; which we use to develop analytically tractable analyses of Pauli tomography. The main result of our work is the demonstration of a deep connection between general QST protocols using IC-bases, and the GUE. This connection in turn enables analytically tractable studies of one-point and two-point functions of the empirical spectral function of tomographically reconstructed states. Essentially, for tomographic protocols yielding normally distributed excess coefficients $\vec{a} \sim \mathcal{N}(0,\Sigma)$, where $\Delta\hat\rho = \mathcal{A}_E(\vec{a}) = \hat\rho-\hat\rho^\prime$ is the error in the tomographic reconstruction, we prove three powerful theorems: \cref{tm:vanishing}, \cref{tm:vanishing-twopoint}, and \cref{tm:vanishing-nondiag}. These theorems, together with the definition of the GUE measure (see \cref{eq:gue}), collectively imply that for any protocol in which $D\mathrm{var}[v]/\overline{v}$ vanishes, and where one is calculating functions of the one-point or two-point empirical spectral distributions, one can substitute samples from the GUE for samples of $\Delta\hat\rho$; where the $\{v_j\}$ are the eigenvalues of $\Sigma$.

We use this flexible and generic result to derive simple analytic forms for expected values and variances of the QST error as quantified by the trace distance $\mathbb{E}[\|\Delta\hat\rho\|_\mathrm{Tr}]$ and $\mathrm{var}[\|\Delta\hat\rho\|_\mathrm{Tr}]$ in \cref{sec:errors}. We validate these analytic forms numerically for both a na\"ive tomographic protocol, and a version of QWC (a more sophisticated protocol which takes advantage of simultaneous measurements) in \cref{sec:numerics}. In \cref{sec:complexity} we reproduce a recently developed non-adaptive Pauli tomography sample complexity bound and find it to be tight for the QWC protocol we consider - supporting the conjecture of Ref.~\cite{acharya2025pauli} that said bound is tight. Finally, in \cref{sec:rephysicalization}, we demonstrate that said bound does not change under application of the most widely-used rephysicalization procedure. The variety, and simplicity, of these calculations demonstrates the flexibility and broad applicability of our RMT approach; and lays the foundation for more comprehensive studies of broad classes of QST protocols from an RMT perspective in the future.

\section{Acknowledgements}
\label{sec:acknowledgements}

A.N.-K. thanks C. Ortega-Taberner for fruitful conversations surrounding the evaluation of the variance integral. J.G. is supported by an SFI-Royal Society University Research Fellowship and is grateful to IBM Ireland and Microsoft Ireland for generous financial support. N.K. thanks S. Zhuk and N. Mariella for helpful discussions.

\section{Author Contributions}
A.N.-K. and N.K. both conceptualised this work, carried out numerical experiments, and wrote the manuscript. N.K. developed the bulk of the mathematical formalisation. All authors contributed to reviewing and editing the manuscript.

\bibliography{refs}

\begin{thebibliography}{55}%
\makeatletter
\providecommand \@ifxundefined [1]{%
 \@ifx{#1\undefined}
}%
\providecommand \@ifnum [1]{%
 \ifnum #1\expandafter \@firstoftwo
 \else \expandafter \@secondoftwo
 \fi
}%
\providecommand \@ifx [1]{%
 \ifx #1\expandafter \@firstoftwo
 \else \expandafter \@secondoftwo
 \fi
}%
\providecommand \natexlab [1]{#1}%
\providecommand \enquote  [1]{``#1''}%
\providecommand \bibnamefont  [1]{#1}%
\providecommand \bibfnamefont [1]{#1}%
\providecommand \citenamefont [1]{#1}%
\providecommand \href@noop [0]{\@secondoftwo}%
\providecommand \href [0]{\begingroup \@sanitize@url \@href}%
\providecommand \@href[1]{\@@startlink{#1}\@@href}%
\providecommand \@@href[1]{\endgroup#1\@@endlink}%
\providecommand \@sanitize@url [0]{\catcode `\\12\catcode `\$12\catcode `\&12\catcode `\#12\catcode `\^12\catcode `\_12\catcode `\%12\relax}%
\providecommand \@@startlink[1]{}%
\providecommand \@@endlink[0]{}%
\providecommand \url  [0]{\begingroup\@sanitize@url \@url }%
\providecommand \@url [1]{\endgroup\@href {#1}{\urlprefix }}%
\providecommand \urlprefix  [0]{URL }%
\providecommand \Eprint [0]{\href }%
\providecommand \doibase [0]{https://doi.org/}%
\providecommand \selectlanguage [0]{\@gobble}%
\providecommand \bibinfo  [0]{\@secondoftwo}%
\providecommand \bibfield  [0]{\@secondoftwo}%
\providecommand \translation [1]{[#1]}%
\providecommand \BibitemOpen [0]{}%
\providecommand \bibitemStop [0]{}%
\providecommand \bibitemNoStop [0]{.\EOS\space}%
\providecommand \EOS [0]{\spacefactor3000\relax}%
\providecommand \BibitemShut  [1]{\csname bibitem#1\endcsname}%
\let\auto@bib@innerbib\@empty
\bibitem [{\citenamefont {Bagan}\ \emph {et~al.}(2004)\citenamefont {Bagan}, \citenamefont {Baig}, \citenamefont {Munoz-Tapia},\ and\ \citenamefont {Rodriguez}}]{bagan2004collective}%
  \BibitemOpen
  \bibfield  {author} {\bibinfo {author} {\bibfnamefont {E.}~\bibnamefont {Bagan}}, \bibinfo {author} {\bibfnamefont {M.}~\bibnamefont {Baig}}, \bibinfo {author} {\bibfnamefont {R.}~\bibnamefont {Munoz-Tapia}},\ and\ \bibinfo {author} {\bibfnamefont {A.}~\bibnamefont {Rodriguez}},\ }\bibfield  {title} {\bibinfo {title} {Collective versus local measurements in a qubit mixed-state estimation},\ }\href@noop {} {\bibfield  {journal} {\bibinfo  {journal} {Physical Review A}\ }\textbf {\bibinfo {volume} {69}},\ \bibinfo {pages} {010304} (\bibinfo {year} {2004})}\BibitemShut {NoStop}%
\bibitem [{\citenamefont {Cai}\ \emph {et~al.}(2016)\citenamefont {Cai}, \citenamefont {Kim}, \citenamefont {Wang}, \citenamefont {Yuan},\ and\ \citenamefont {Zhou}}]{cai2016optimal}%
  \BibitemOpen
  \bibfield  {author} {\bibinfo {author} {\bibfnamefont {T.}~\bibnamefont {Cai}}, \bibinfo {author} {\bibfnamefont {D.}~\bibnamefont {Kim}}, \bibinfo {author} {\bibfnamefont {Y.}~\bibnamefont {Wang}}, \bibinfo {author} {\bibfnamefont {M.}~\bibnamefont {Yuan}},\ and\ \bibinfo {author} {\bibfnamefont {H.~H.}\ \bibnamefont {Zhou}},\ }\bibfield  {title} {\bibinfo {title} {{Optimal large-scale quantum state tomography with Pauli measurements.}},\ }\href {https://doi.org/10.1214/15-AOS1382} {\bibfield  {journal} {\bibinfo  {journal} {Annals Statist.}\ }\textbf {\bibinfo {volume} {44}},\ \bibinfo {pages} {682} (\bibinfo {year} {2016})}\BibitemShut {NoStop}%
\bibitem [{\citenamefont {Keyl}(2006)}]{keyl2006quantum}%
  \BibitemOpen
  \bibfield  {author} {\bibinfo {author} {\bibfnamefont {M.}~\bibnamefont {Keyl}},\ }\bibfield  {title} {\bibinfo {title} {Quantum state estimation and large deviations},\ }\href@noop {} {\bibfield  {journal} {\bibinfo  {journal} {Reviews in Mathematical Physics}\ }\textbf {\bibinfo {volume} {18}},\ \bibinfo {pages} {19} (\bibinfo {year} {2006})}\BibitemShut {NoStop}%
\bibitem [{\citenamefont {Gu{\c{t}}{\u{a}}}\ \emph {et~al.}(2008)\citenamefont {Gu{\c{t}}{\u{a}}}, \citenamefont {Janssens},\ and\ \citenamefont {Kahn}}]{guctua2008optimal}%
  \BibitemOpen
  \bibfield  {author} {\bibinfo {author} {\bibfnamefont {M.}~\bibnamefont {Gu{\c{t}}{\u{a}}}}, \bibinfo {author} {\bibfnamefont {B.}~\bibnamefont {Janssens}},\ and\ \bibinfo {author} {\bibfnamefont {J.}~\bibnamefont {Kahn}},\ }\bibfield  {title} {\bibinfo {title} {Optimal estimation of qubit states with continuous time measurements},\ }\href@noop {} {\bibfield  {journal} {\bibinfo  {journal} {Communications in Mathematical Physics}\ }\textbf {\bibinfo {volume} {277}},\ \bibinfo {pages} {127} (\bibinfo {year} {2008})}\BibitemShut {NoStop}%
\bibitem [{\citenamefont {Nico-Katz}\ \emph {et~al.}(2024)\citenamefont {Nico-Katz}, \citenamefont {Keenan},\ and\ \citenamefont {Goold}}]{nico2024can}%
  \BibitemOpen
  \bibfield  {author} {\bibinfo {author} {\bibfnamefont {A.}~\bibnamefont {Nico-Katz}}, \bibinfo {author} {\bibfnamefont {N.}~\bibnamefont {Keenan}},\ and\ \bibinfo {author} {\bibfnamefont {J.}~\bibnamefont {Goold}},\ }\bibfield  {title} {\bibinfo {title} {Can quantum computers do nothing?},\ }\href@noop {} {\bibfield  {journal} {\bibinfo  {journal} {npj Quantum Information}\ }\textbf {\bibinfo {volume} {10}},\ \bibinfo {pages} {124} (\bibinfo {year} {2024})}\BibitemShut {NoStop}%
\bibitem [{\citenamefont {Yu}(2023)}]{yu2023almost}%
  \BibitemOpen
  \bibfield  {author} {\bibinfo {author} {\bibfnamefont {N.}~\bibnamefont {Yu}},\ }\bibfield  {title} {\bibinfo {title} {Almost tight sample complexity analysis of quantum identity testing by pauli measurements},\ }\href@noop {} {\bibfield  {journal} {\bibinfo  {journal} {IEEE Transactions on Information Theory}\ }\textbf {\bibinfo {volume} {69}},\ \bibinfo {pages} {5060} (\bibinfo {year} {2023})}\BibitemShut {NoStop}%
\bibitem [{\citenamefont {Huang}\ \emph {et~al.}(2024)\citenamefont {Huang}, \citenamefont {Liu}, \citenamefont {Broughton}, \citenamefont {Kim}, \citenamefont {Anshu}, \citenamefont {Landau},\ and\ \citenamefont {McClean}}]{huang2024learning}%
  \BibitemOpen
  \bibfield  {author} {\bibinfo {author} {\bibfnamefont {H.-Y.}\ \bibnamefont {Huang}}, \bibinfo {author} {\bibfnamefont {Y.}~\bibnamefont {Liu}}, \bibinfo {author} {\bibfnamefont {M.}~\bibnamefont {Broughton}}, \bibinfo {author} {\bibfnamefont {I.}~\bibnamefont {Kim}}, \bibinfo {author} {\bibfnamefont {A.}~\bibnamefont {Anshu}}, \bibinfo {author} {\bibfnamefont {Z.}~\bibnamefont {Landau}},\ and\ \bibinfo {author} {\bibfnamefont {J.~R.}\ \bibnamefont {McClean}},\ }\bibfield  {title} {\bibinfo {title} {Learning shallow quantum circuits},\ }in\ \href@noop {} {\emph {\bibinfo {booktitle} {Proceedings of the 56th Annual ACM Symposium on Theory of Computing}}}\ (\bibinfo {year} {2024})\ pp.\ \bibinfo {pages} {1343--1351}\BibitemShut {NoStop}%
\bibitem [{\citenamefont {Flammia}\ \emph {et~al.}(2012)\citenamefont {Flammia}, \citenamefont {Gross}, \citenamefont {Liu},\ and\ \citenamefont {Eisert}}]{flammia2012quantum}%
  \BibitemOpen
  \bibfield  {author} {\bibinfo {author} {\bibfnamefont {S.~T.}\ \bibnamefont {Flammia}}, \bibinfo {author} {\bibfnamefont {D.}~\bibnamefont {Gross}}, \bibinfo {author} {\bibfnamefont {Y.-K.}\ \bibnamefont {Liu}},\ and\ \bibinfo {author} {\bibfnamefont {J.}~\bibnamefont {Eisert}},\ }\bibfield  {title} {\bibinfo {title} {Quantum tomography via compressed sensing: error bounds, sample complexity and efficient estimators},\ }\href@noop {} {\bibfield  {journal} {\bibinfo  {journal} {New Journal of Physics}\ }\textbf {\bibinfo {volume} {14}},\ \bibinfo {pages} {095022} (\bibinfo {year} {2012})}\BibitemShut {NoStop}%
\bibitem [{\citenamefont {Lowe}\ and\ \citenamefont {Nayak}(2022)}]{lowe2022lower}%
  \BibitemOpen
  \bibfield  {author} {\bibinfo {author} {\bibfnamefont {A.}~\bibnamefont {Lowe}}\ and\ \bibinfo {author} {\bibfnamefont {A.}~\bibnamefont {Nayak}},\ }\bibfield  {title} {\bibinfo {title} {Lower bounds for learning quantum states with single-copy measurements},\ }\href@noop {} {\bibfield  {journal} {\bibinfo  {journal} {arXiv preprint arXiv:2207.14438}\ } (\bibinfo {year} {2022})}\BibitemShut {NoStop}%
\bibitem [{\citenamefont {Yuen}(2023)}]{yuen2023improved}%
  \BibitemOpen
  \bibfield  {author} {\bibinfo {author} {\bibfnamefont {H.}~\bibnamefont {Yuen}},\ }\bibfield  {title} {\bibinfo {title} {An improved sample complexity lower bound for (fidelity) quantum state tomography},\ }\href@noop {} {\bibfield  {journal} {\bibinfo  {journal} {Quantum}\ }\textbf {\bibinfo {volume} {7}},\ \bibinfo {pages} {890} (\bibinfo {year} {2023})}\BibitemShut {NoStop}%
\bibitem [{\citenamefont {Bubeck}\ \emph {et~al.}(2020)\citenamefont {Bubeck}, \citenamefont {Chen},\ and\ \citenamefont {Li}}]{bubeck2020entanglement}%
  \BibitemOpen
  \bibfield  {author} {\bibinfo {author} {\bibfnamefont {S.}~\bibnamefont {Bubeck}}, \bibinfo {author} {\bibfnamefont {S.}~\bibnamefont {Chen}},\ and\ \bibinfo {author} {\bibfnamefont {J.}~\bibnamefont {Li}},\ }\bibfield  {title} {\bibinfo {title} {Entanglement is necessary for optimal quantum property testing},\ }in\ \href@noop {} {\emph {\bibinfo {booktitle} {2020 IEEE 61st Annual Symposium on Foundations of Computer Science (FOCS)}}}\ (\bibinfo {organization} {IEEE},\ \bibinfo {year} {2020})\ pp.\ \bibinfo {pages} {692--703}\BibitemShut {NoStop}%
\bibitem [{\citenamefont {O'Donnell}\ and\ \citenamefont {Wright}(2016)}]{o2016efficient}%
  \BibitemOpen
  \bibfield  {author} {\bibinfo {author} {\bibfnamefont {R.}~\bibnamefont {O'Donnell}}\ and\ \bibinfo {author} {\bibfnamefont {J.}~\bibnamefont {Wright}},\ }\bibfield  {title} {\bibinfo {title} {Efficient quantum tomography},\ }in\ \href@noop {} {\emph {\bibinfo {booktitle} {Proceedings of the forty-eighth annual ACM symposium on Theory of Computing}}}\ (\bibinfo {year} {2016})\ pp.\ \bibinfo {pages} {899--912}\BibitemShut {NoStop}%
\bibitem [{\citenamefont {O'Donnell}\ and\ \citenamefont {Wright}(2017)}]{o2017efficient}%
  \BibitemOpen
  \bibfield  {author} {\bibinfo {author} {\bibfnamefont {R.}~\bibnamefont {O'Donnell}}\ and\ \bibinfo {author} {\bibfnamefont {J.}~\bibnamefont {Wright}},\ }\bibfield  {title} {\bibinfo {title} {Efficient quantum tomography ii},\ }in\ \href@noop {} {\emph {\bibinfo {booktitle} {Proceedings of the 49th Annual ACM SIGACT Symposium on Theory of Computing}}}\ (\bibinfo {year} {2017})\ pp.\ \bibinfo {pages} {962--974}\BibitemShut {NoStop}%
\bibitem [{\citenamefont {Haah}\ \emph {et~al.}(2016)\citenamefont {Haah}, \citenamefont {Harrow}, \citenamefont {Ji}, \citenamefont {Wu},\ and\ \citenamefont {Yu}}]{haah2016sample}%
  \BibitemOpen
  \bibfield  {author} {\bibinfo {author} {\bibfnamefont {J.}~\bibnamefont {Haah}}, \bibinfo {author} {\bibfnamefont {A.~W.}\ \bibnamefont {Harrow}}, \bibinfo {author} {\bibfnamefont {Z.}~\bibnamefont {Ji}}, \bibinfo {author} {\bibfnamefont {X.}~\bibnamefont {Wu}},\ and\ \bibinfo {author} {\bibfnamefont {N.}~\bibnamefont {Yu}},\ }\bibfield  {title} {\bibinfo {title} {Sample-optimal tomography of quantum states},\ }in\ \href@noop {} {\emph {\bibinfo {booktitle} {Proceedings of the forty-eighth annual ACM symposium on Theory of Computing}}}\ (\bibinfo {year} {2016})\ pp.\ \bibinfo {pages} {913--925}\BibitemShut {NoStop}%
\bibitem [{\citenamefont {Chen}\ \emph {et~al.}(2024)\citenamefont {Chen}, \citenamefont {Li},\ and\ \citenamefont {Liu}}]{chen2024optimal}%
  \BibitemOpen
  \bibfield  {author} {\bibinfo {author} {\bibfnamefont {S.}~\bibnamefont {Chen}}, \bibinfo {author} {\bibfnamefont {J.}~\bibnamefont {Li}},\ and\ \bibinfo {author} {\bibfnamefont {A.}~\bibnamefont {Liu}},\ }\bibfield  {title} {\bibinfo {title} {An optimal tradeoff between entanglement and copy complexity for state tomography},\ }in\ \href@noop {} {\emph {\bibinfo {booktitle} {Proceedings of the 56th Annual ACM Symposium on Theory of Computing}}}\ (\bibinfo {year} {2024})\ pp.\ \bibinfo {pages} {1331--1342}\BibitemShut {NoStop}%
\bibitem [{\citenamefont {Flammia}\ and\ \citenamefont {O'Donnell}(2024)}]{flammia2024quantum}%
  \BibitemOpen
  \bibfield  {author} {\bibinfo {author} {\bibfnamefont {S.~T.}\ \bibnamefont {Flammia}}\ and\ \bibinfo {author} {\bibfnamefont {R.}~\bibnamefont {O'Donnell}},\ }\bibfield  {title} {\bibinfo {title} {Quantum chi-squared tomography and mutual information testing},\ }\href@noop {} {\bibfield  {journal} {\bibinfo  {journal} {Quantum}\ }\textbf {\bibinfo {volume} {8}},\ \bibinfo {pages} {1381} (\bibinfo {year} {2024})}\BibitemShut {NoStop}%
\bibitem [{\citenamefont {Akhtar}\ \emph {et~al.}(2022)\citenamefont {Akhtar}, \citenamefont {Hu},\ and\ \citenamefont {You}}]{akhtar2209scalable}%
  \BibitemOpen
  \bibfield  {author} {\bibinfo {author} {\bibfnamefont {A.~A.}\ \bibnamefont {Akhtar}}, \bibinfo {author} {\bibfnamefont {H.-Y.}\ \bibnamefont {Hu}},\ and\ \bibinfo {author} {\bibfnamefont {Y.-Z.}\ \bibnamefont {You}},\ }\bibfield  {title} {\bibinfo {title} {Scalable and flexible classical shadow tomography with tensor networks},\ }\href@noop {} {\bibfield  {journal} {\bibinfo  {journal} {arXiv preprint arXiv:2209.02093}\ } (\bibinfo {year} {2022})}\BibitemShut {NoStop}%
\bibitem [{\citenamefont {Zhang}\ \emph {et~al.}(2024)\citenamefont {Zhang}, \citenamefont {Smith}, \citenamefont {Smolin}, \citenamefont {Liu}, \citenamefont {Peng}, \citenamefont {Zhao}, \citenamefont {Girolami}, \citenamefont {Ma}, \citenamefont {Yuan},\ and\ \citenamefont {Lu}}]{zhang2024quantification}%
  \BibitemOpen
  \bibfield  {author} {\bibinfo {author} {\bibfnamefont {T.}~\bibnamefont {Zhang}}, \bibinfo {author} {\bibfnamefont {G.}~\bibnamefont {Smith}}, \bibinfo {author} {\bibfnamefont {J.~A.}\ \bibnamefont {Smolin}}, \bibinfo {author} {\bibfnamefont {L.}~\bibnamefont {Liu}}, \bibinfo {author} {\bibfnamefont {X.-J.}\ \bibnamefont {Peng}}, \bibinfo {author} {\bibfnamefont {Q.}~\bibnamefont {Zhao}}, \bibinfo {author} {\bibfnamefont {D.}~\bibnamefont {Girolami}}, \bibinfo {author} {\bibfnamefont {X.}~\bibnamefont {Ma}}, \bibinfo {author} {\bibfnamefont {X.}~\bibnamefont {Yuan}},\ and\ \bibinfo {author} {\bibfnamefont {H.}~\bibnamefont {Lu}},\ }\bibfield  {title} {\bibinfo {title} {Quantification of entanglement and coherence with purity detection},\ }\href@noop {} {\bibfield  {journal} {\bibinfo  {journal} {npj Quantum Information}\ }\textbf {\bibinfo {volume} {10}},\ \bibinfo {pages} {60} (\bibinfo {year} {2024})}\BibitemShut {NoStop}%
\bibitem [{\citenamefont {Sengupta}\ \emph {et~al.}(2025)\citenamefont {Sengupta}, \citenamefont {Chatterjee}, \citenamefont {Sreejith},\ and\ \citenamefont {Mahesh}}]{sengupta2025partial}%
  \BibitemOpen
  \bibfield  {author} {\bibinfo {author} {\bibfnamefont {A.}~\bibnamefont {Sengupta}}, \bibinfo {author} {\bibfnamefont {A.}~\bibnamefont {Chatterjee}}, \bibinfo {author} {\bibfnamefont {G.}~\bibnamefont {Sreejith}},\ and\ \bibinfo {author} {\bibfnamefont {T.}~\bibnamefont {Mahesh}},\ }\bibfield  {title} {\bibinfo {title} {Partial quantum shadow tomography for structured operators and its experimental demonstration using nmr},\ }\href@noop {} {\bibfield  {journal} {\bibinfo  {journal} {arXiv preprint arXiv:2503.14491}\ } (\bibinfo {year} {2025})}\BibitemShut {NoStop}%
\bibitem [{\citenamefont {Rath}\ \emph {et~al.}(2021)\citenamefont {Rath}, \citenamefont {Branciard}, \citenamefont {Minguzzi},\ and\ \citenamefont {Vermersch}}]{rath2021quantum}%
  \BibitemOpen
  \bibfield  {author} {\bibinfo {author} {\bibfnamefont {A.}~\bibnamefont {Rath}}, \bibinfo {author} {\bibfnamefont {C.}~\bibnamefont {Branciard}}, \bibinfo {author} {\bibfnamefont {A.}~\bibnamefont {Minguzzi}},\ and\ \bibinfo {author} {\bibfnamefont {B.}~\bibnamefont {Vermersch}},\ }\bibfield  {title} {\bibinfo {title} {Quantum fisher information from randomized measurements},\ }\href@noop {} {\bibfield  {journal} {\bibinfo  {journal} {Physical Review Letters}\ }\textbf {\bibinfo {volume} {127}},\ \bibinfo {pages} {260501} (\bibinfo {year} {2021})}\BibitemShut {NoStop}%
\bibitem [{\citenamefont {Aaronson}(2018)}]{aaronson2018shadow}%
  \BibitemOpen
  \bibfield  {author} {\bibinfo {author} {\bibfnamefont {S.}~\bibnamefont {Aaronson}},\ }\bibfield  {title} {\bibinfo {title} {Shadow tomography of quantum states},\ }in\ \href@noop {} {\emph {\bibinfo {booktitle} {Proceedings of the 50th annual ACM SIGACT symposium on theory of computing}}}\ (\bibinfo {year} {2018})\ pp.\ \bibinfo {pages} {325--338}\BibitemShut {NoStop}%
\bibitem [{\citenamefont {Elben}\ \emph {et~al.}(2023)\citenamefont {Elben}, \citenamefont {Flammia}, \citenamefont {Huang}, \citenamefont {Kueng}, \citenamefont {Preskill}, \citenamefont {Vermersch},\ and\ \citenamefont {Zoller}}]{elben2023randomized}%
  \BibitemOpen
  \bibfield  {author} {\bibinfo {author} {\bibfnamefont {A.}~\bibnamefont {Elben}}, \bibinfo {author} {\bibfnamefont {S.~T.}\ \bibnamefont {Flammia}}, \bibinfo {author} {\bibfnamefont {H.-Y.}\ \bibnamefont {Huang}}, \bibinfo {author} {\bibfnamefont {R.}~\bibnamefont {Kueng}}, \bibinfo {author} {\bibfnamefont {J.}~\bibnamefont {Preskill}}, \bibinfo {author} {\bibfnamefont {B.}~\bibnamefont {Vermersch}},\ and\ \bibinfo {author} {\bibfnamefont {P.}~\bibnamefont {Zoller}},\ }\bibfield  {title} {\bibinfo {title} {The randomized measurement toolbox},\ }\href@noop {} {\bibfield  {journal} {\bibinfo  {journal} {Nature Reviews Physics}\ }\textbf {\bibinfo {volume} {5}},\ \bibinfo {pages} {9} (\bibinfo {year} {2023})}\BibitemShut {NoStop}%
\bibitem [{\citenamefont {Chen}\ \emph {et~al.}(2022{\natexlab{a}})\citenamefont {Chen}, \citenamefont {Cotler}, \citenamefont {Huang},\ and\ \citenamefont {Li}}]{chen2022exponential}%
  \BibitemOpen
  \bibfield  {author} {\bibinfo {author} {\bibfnamefont {S.}~\bibnamefont {Chen}}, \bibinfo {author} {\bibfnamefont {J.}~\bibnamefont {Cotler}}, \bibinfo {author} {\bibfnamefont {H.-Y.}\ \bibnamefont {Huang}},\ and\ \bibinfo {author} {\bibfnamefont {J.}~\bibnamefont {Li}},\ }\bibfield  {title} {\bibinfo {title} {Exponential separations between learning with and without quantum memory},\ }in\ \href@noop {} {\emph {\bibinfo {booktitle} {2021 IEEE 62nd Annual Symposium on Foundations of Computer Science (FOCS)}}}\ (\bibinfo {organization} {IEEE},\ \bibinfo {year} {2022})\ pp.\ \bibinfo {pages} {574--585}\BibitemShut {NoStop}%
\bibitem [{\citenamefont {Huang}\ \emph {et~al.}(2020)\citenamefont {Huang}, \citenamefont {Kueng},\ and\ \citenamefont {Preskill}}]{huang2020predicting}%
  \BibitemOpen
  \bibfield  {author} {\bibinfo {author} {\bibfnamefont {H.-Y.}\ \bibnamefont {Huang}}, \bibinfo {author} {\bibfnamefont {R.}~\bibnamefont {Kueng}},\ and\ \bibinfo {author} {\bibfnamefont {J.}~\bibnamefont {Preskill}},\ }\bibfield  {title} {\bibinfo {title} {Predicting many properties of a quantum system from very few measurements},\ }\href@noop {} {\bibfield  {journal} {\bibinfo  {journal} {Nature Physics}\ }\textbf {\bibinfo {volume} {16}},\ \bibinfo {pages} {1050} (\bibinfo {year} {2020})}\BibitemShut {NoStop}%
\bibitem [{\citenamefont {Garc{\'\i}a-P{\'e}rez}\ \emph {et~al.}(2020)\citenamefont {Garc{\'\i}a-P{\'e}rez}, \citenamefont {Rossi}, \citenamefont {Sokolov}, \citenamefont {Borrelli},\ and\ \citenamefont {Maniscalco}}]{garcia2020pairwise}%
  \BibitemOpen
  \bibfield  {author} {\bibinfo {author} {\bibfnamefont {G.}~\bibnamefont {Garc{\'\i}a-P{\'e}rez}}, \bibinfo {author} {\bibfnamefont {M.~A.}\ \bibnamefont {Rossi}}, \bibinfo {author} {\bibfnamefont {B.}~\bibnamefont {Sokolov}}, \bibinfo {author} {\bibfnamefont {E.-M.}\ \bibnamefont {Borrelli}},\ and\ \bibinfo {author} {\bibfnamefont {S.}~\bibnamefont {Maniscalco}},\ }\bibfield  {title} {\bibinfo {title} {Pairwise tomography networks for many-body quantum systems},\ }\href@noop {} {\bibfield  {journal} {\bibinfo  {journal} {Physical Review Research}\ }\textbf {\bibinfo {volume} {2}},\ \bibinfo {pages} {023393} (\bibinfo {year} {2020})}\BibitemShut {NoStop}%
\bibitem [{\citenamefont {Cotler}\ and\ \citenamefont {Wilczek}(2020)}]{cotler2020quantum}%
  \BibitemOpen
  \bibfield  {author} {\bibinfo {author} {\bibfnamefont {J.}~\bibnamefont {Cotler}}\ and\ \bibinfo {author} {\bibfnamefont {F.}~\bibnamefont {Wilczek}},\ }\bibfield  {title} {\bibinfo {title} {Quantum overlapping tomography},\ }\href@noop {} {\bibfield  {journal} {\bibinfo  {journal} {Physical review letters}\ }\textbf {\bibinfo {volume} {124}},\ \bibinfo {pages} {100401} (\bibinfo {year} {2020})}\BibitemShut {NoStop}%
\bibitem [{\citenamefont {B{\u{a}}descu}\ \emph {et~al.}(2019)\citenamefont {B{\u{a}}descu}, \citenamefont {O'Donnell},\ and\ \citenamefont {Wright}}]{buadescu2019quantum}%
  \BibitemOpen
  \bibfield  {author} {\bibinfo {author} {\bibfnamefont {C.}~\bibnamefont {B{\u{a}}descu}}, \bibinfo {author} {\bibfnamefont {R.}~\bibnamefont {O'Donnell}},\ and\ \bibinfo {author} {\bibfnamefont {J.}~\bibnamefont {Wright}},\ }\bibfield  {title} {\bibinfo {title} {Quantum state certification},\ }in\ \href@noop {} {\emph {\bibinfo {booktitle} {Proceedings of the 51st Annual ACM SIGACT Symposium on Theory of Computing}}}\ (\bibinfo {year} {2019})\ pp.\ \bibinfo {pages} {503--514}\BibitemShut {NoStop}%
\bibitem [{\citenamefont {Chen}\ \emph {et~al.}(2022{\natexlab{b}})\citenamefont {Chen}, \citenamefont {Li}, \citenamefont {Huang},\ and\ \citenamefont {Liu}}]{chen2022tight}%
  \BibitemOpen
  \bibfield  {author} {\bibinfo {author} {\bibfnamefont {S.}~\bibnamefont {Chen}}, \bibinfo {author} {\bibfnamefont {J.}~\bibnamefont {Li}}, \bibinfo {author} {\bibfnamefont {B.}~\bibnamefont {Huang}},\ and\ \bibinfo {author} {\bibfnamefont {A.}~\bibnamefont {Liu}},\ }\bibfield  {title} {\bibinfo {title} {Tight bounds for quantum state certification with incoherent measurements},\ }in\ \href@noop {} {\emph {\bibinfo {booktitle} {2022 IEEE 63rd Annual Symposium on Foundations of Computer Science (FOCS)}}}\ (\bibinfo {organization} {IEEE},\ \bibinfo {year} {2022})\ pp.\ \bibinfo {pages} {1205--1213}\BibitemShut {NoStop}%
\bibitem [{\citenamefont {Liu}\ and\ \citenamefont {Acharya}(2024)}]{liu2024role}%
  \BibitemOpen
  \bibfield  {author} {\bibinfo {author} {\bibfnamefont {Y.}~\bibnamefont {Liu}}\ and\ \bibinfo {author} {\bibfnamefont {J.}~\bibnamefont {Acharya}},\ }\bibfield  {title} {\bibinfo {title} {The role of randomness in quantum state certification with unentangled measurements},\ }in\ \href@noop {} {\emph {\bibinfo {booktitle} {The Thirty Seventh Annual Conference on Learning Theory}}}\ (\bibinfo {organization} {PMLR},\ \bibinfo {year} {2024})\ pp.\ \bibinfo {pages} {3523--3555}\BibitemShut {NoStop}%
\bibitem [{\citenamefont {Gu{\c{t}}{\u{a}}}\ \emph {et~al.}(2020)\citenamefont {Gu{\c{t}}{\u{a}}}, \citenamefont {Kahn}, \citenamefont {Kueng},\ and\ \citenamefont {Tropp}}]{guctua2020fast}%
  \BibitemOpen
  \bibfield  {author} {\bibinfo {author} {\bibfnamefont {M.}~\bibnamefont {Gu{\c{t}}{\u{a}}}}, \bibinfo {author} {\bibfnamefont {J.}~\bibnamefont {Kahn}}, \bibinfo {author} {\bibfnamefont {R.}~\bibnamefont {Kueng}},\ and\ \bibinfo {author} {\bibfnamefont {J.~A.}\ \bibnamefont {Tropp}},\ }\bibfield  {title} {\bibinfo {title} {Fast state tomography with optimal error bounds},\ }\href@noop {} {\bibfield  {journal} {\bibinfo  {journal} {Journal of Physics A: Mathematical and Theoretical}\ }\textbf {\bibinfo {volume} {53}},\ \bibinfo {pages} {204001} (\bibinfo {year} {2020})}\BibitemShut {NoStop}%
\bibitem [{\citenamefont {Acharya}\ \emph {et~al.}(2025)\citenamefont {Acharya}, \citenamefont {Dharmavarapu}, \citenamefont {Liu},\ and\ \citenamefont {Yu}}]{acharya2025pauli}%
  \BibitemOpen
  \bibfield  {author} {\bibinfo {author} {\bibfnamefont {J.}~\bibnamefont {Acharya}}, \bibinfo {author} {\bibfnamefont {A.}~\bibnamefont {Dharmavarapu}}, \bibinfo {author} {\bibfnamefont {Y.}~\bibnamefont {Liu}},\ and\ \bibinfo {author} {\bibfnamefont {N.}~\bibnamefont {Yu}},\ }\bibfield  {title} {\bibinfo {title} {Pauli measurements are not optimal for single-copy tomography},\ }\href@noop {} {\bibfield  {journal} {\bibinfo  {journal} {arXiv preprint arXiv:2502.18170}\ } (\bibinfo {year} {2025})}\BibitemShut {NoStop}%
\bibitem [{\citenamefont {Gross}\ \emph {et~al.}(2010)\citenamefont {Gross}, \citenamefont {Liu}, \citenamefont {Flammia}, \citenamefont {Becker},\ and\ \citenamefont {Eisert}}]{gross2010quantum}%
  \BibitemOpen
  \bibfield  {author} {\bibinfo {author} {\bibfnamefont {D.}~\bibnamefont {Gross}}, \bibinfo {author} {\bibfnamefont {Y.-K.}\ \bibnamefont {Liu}}, \bibinfo {author} {\bibfnamefont {S.~T.}\ \bibnamefont {Flammia}}, \bibinfo {author} {\bibfnamefont {S.}~\bibnamefont {Becker}},\ and\ \bibinfo {author} {\bibfnamefont {J.}~\bibnamefont {Eisert}},\ }\bibfield  {title} {\bibinfo {title} {Quantum state tomography via compressed sensing},\ }\href@noop {} {\bibfield  {journal} {\bibinfo  {journal} {Physical review letters}\ }\textbf {\bibinfo {volume} {105}},\ \bibinfo {pages} {150401} (\bibinfo {year} {2010})}\BibitemShut {NoStop}%
\bibitem [{\citenamefont {Mahler}\ \emph {et~al.}(2013)\citenamefont {Mahler}, \citenamefont {Rozema}, \citenamefont {Darabi}, \citenamefont {Ferrie}, \citenamefont {Blume-Kohout},\ and\ \citenamefont {Steinberg}}]{mahler2013adaptive}%
  \BibitemOpen
  \bibfield  {author} {\bibinfo {author} {\bibfnamefont {D.~H.}\ \bibnamefont {Mahler}}, \bibinfo {author} {\bibfnamefont {L.~A.}\ \bibnamefont {Rozema}}, \bibinfo {author} {\bibfnamefont {A.}~\bibnamefont {Darabi}}, \bibinfo {author} {\bibfnamefont {C.}~\bibnamefont {Ferrie}}, \bibinfo {author} {\bibfnamefont {R.}~\bibnamefont {Blume-Kohout}},\ and\ \bibinfo {author} {\bibfnamefont {A.~M.}\ \bibnamefont {Steinberg}},\ }\bibfield  {title} {\bibinfo {title} {Adaptive quantum state tomography improves accuracy quadratically},\ }\href@noop {} {\bibfield  {journal} {\bibinfo  {journal} {Physical review letters}\ }\textbf {\bibinfo {volume} {111}},\ \bibinfo {pages} {183601} (\bibinfo {year} {2013})}\BibitemShut {NoStop}%
\bibitem [{\citenamefont {Chen}\ \emph {et~al.}(2023)\citenamefont {Chen}, \citenamefont {Huang}, \citenamefont {Li}, \citenamefont {Liu},\ and\ \citenamefont {Sellke}}]{chen2023does}%
  \BibitemOpen
  \bibfield  {author} {\bibinfo {author} {\bibfnamefont {S.}~\bibnamefont {Chen}}, \bibinfo {author} {\bibfnamefont {B.}~\bibnamefont {Huang}}, \bibinfo {author} {\bibfnamefont {J.}~\bibnamefont {Li}}, \bibinfo {author} {\bibfnamefont {A.}~\bibnamefont {Liu}},\ and\ \bibinfo {author} {\bibfnamefont {M.}~\bibnamefont {Sellke}},\ }\bibfield  {title} {\bibinfo {title} {When does adaptivity help for quantum state learning?},\ }in\ \href@noop {} {\emph {\bibinfo {booktitle} {2023 IEEE 64th Annual Symposium on Foundations of Computer Science (FOCS)}}}\ (\bibinfo {organization} {IEEE},\ \bibinfo {year} {2023})\ pp.\ \bibinfo {pages} {391--404}\BibitemShut {NoStop}%
\bibitem [{Note1()}]{Note1}%
  \BibitemOpen
  \bibinfo {note} {Non-adaptive QST protocols are those in which the decision on which POVM element $P_j \in \protect \mathcal {P}_N$ to measure during a given shot is independent of previous shot outcomes.}\BibitemShut {Stop}%
\bibitem [{\citenamefont {Dietz}(2006)}]{dietz2006generalized}%
  \BibitemOpen
  \bibfield  {author} {\bibinfo {author} {\bibfnamefont {K.}~\bibnamefont {Dietz}},\ }\bibfield  {title} {\bibinfo {title} {Generalized bloch spheres for m-qubit states},\ }\href@noop {} {\bibfield  {journal} {\bibinfo  {journal} {Journal of Physics A: Mathematical and General}\ }\textbf {\bibinfo {volume} {39}},\ \bibinfo {pages} {1433} (\bibinfo {year} {2006})}\BibitemShut {NoStop}%
\bibitem [{\citenamefont {Wie}(2020)}]{wie2020two}%
  \BibitemOpen
  \bibfield  {author} {\bibinfo {author} {\bibfnamefont {C.-R.}\ \bibnamefont {Wie}},\ }\bibfield  {title} {\bibinfo {title} {Two-qubit bloch sphere},\ }\href@noop {} {\bibfield  {journal} {\bibinfo  {journal} {Physics}\ }\textbf {\bibinfo {volume} {2}},\ \bibinfo {pages} {383} (\bibinfo {year} {2020})}\BibitemShut {NoStop}%
\bibitem [{\citenamefont {Filatov}\ and\ \citenamefont {Auzinsh}(2024)}]{filatov2024towards}%
  \BibitemOpen
  \bibfield  {author} {\bibinfo {author} {\bibfnamefont {S.}~\bibnamefont {Filatov}}\ and\ \bibinfo {author} {\bibfnamefont {M.}~\bibnamefont {Auzinsh}},\ }\bibfield  {title} {\bibinfo {title} {Towards two bloch sphere representation of pure two-qubit states and unitaries},\ }\href@noop {} {\bibfield  {journal} {\bibinfo  {journal} {Entropy}\ }\textbf {\bibinfo {volume} {26}},\ \bibinfo {pages} {280} (\bibinfo {year} {2024})}\BibitemShut {NoStop}%
\bibitem [{\citenamefont {Singh}\ \emph {et~al.}(2016)\citenamefont {Singh}, \citenamefont {Dorai} \emph {et~al.}}]{singh2016constructing}%
  \BibitemOpen
  \bibfield  {author} {\bibinfo {author} {\bibfnamefont {H.}~\bibnamefont {Singh}}, \bibinfo {author} {\bibfnamefont {K.}~\bibnamefont {Dorai}}, \emph {et~al.},\ }\bibfield  {title} {\bibinfo {title} {Constructing valid density matrices on an nmr quantum information processor via maximum likelihood estimation},\ }\href@noop {} {\bibfield  {journal} {\bibinfo  {journal} {Physics Letters A}\ }\textbf {\bibinfo {volume} {380}},\ \bibinfo {pages} {3051} (\bibinfo {year} {2016})}\BibitemShut {NoStop}%
\bibitem [{\citenamefont {Smolin}\ \emph {et~al.}(2012)\citenamefont {Smolin}, \citenamefont {Gambetta},\ and\ \citenamefont {Smith}}]{smolin2012efficient}%
  \BibitemOpen
  \bibfield  {author} {\bibinfo {author} {\bibfnamefont {J.~A.}\ \bibnamefont {Smolin}}, \bibinfo {author} {\bibfnamefont {J.~M.}\ \bibnamefont {Gambetta}},\ and\ \bibinfo {author} {\bibfnamefont {G.}~\bibnamefont {Smith}},\ }\bibfield  {title} {\bibinfo {title} {Efficient method for computing the maximum-likelihood quantum state from measurements with additive gaussian noise},\ }\href@noop {} {\bibfield  {journal} {\bibinfo  {journal} {Physical review letters}\ }\textbf {\bibinfo {volume} {108}},\ \bibinfo {pages} {070502} (\bibinfo {year} {2012})}\BibitemShut {NoStop}%
\bibitem [{\citenamefont {Ferrie}\ and\ \citenamefont {Blume-Kohout}(2018)}]{ferrie2018maximum}%
  \BibitemOpen
  \bibfield  {author} {\bibinfo {author} {\bibfnamefont {C.}~\bibnamefont {Ferrie}}\ and\ \bibinfo {author} {\bibfnamefont {R.}~\bibnamefont {Blume-Kohout}},\ }\bibfield  {title} {\bibinfo {title} {Maximum likelihood quantum state tomography is inadmissible},\ }\href@noop {} {\bibfield  {journal} {\bibinfo  {journal} {arXiv preprint arXiv:1808.01072}\ } (\bibinfo {year} {2018})}\BibitemShut {NoStop}%
\bibitem [{\citenamefont {Siddhu}(2019)}]{siddhu2019maximum}%
  \BibitemOpen
  \bibfield  {author} {\bibinfo {author} {\bibfnamefont {V.}~\bibnamefont {Siddhu}},\ }\bibfield  {title} {\bibinfo {title} {Maximum a posteriori probability estimates for quantum tomography},\ }\href@noop {} {\bibfield  {journal} {\bibinfo  {journal} {Physical Review A}\ }\textbf {\bibinfo {volume} {99}},\ \bibinfo {pages} {012342} (\bibinfo {year} {2019})}\BibitemShut {NoStop}%
\bibitem [{\citenamefont {Mehta}(2004)}]{mehta2004random}%
  \BibitemOpen
  \bibfield  {author} {\bibinfo {author} {\bibfnamefont {M.~L.}\ \bibnamefont {Mehta}},\ }\href@noop {} {\emph {\bibinfo {title} {Random matrices}}},\ Vol.\ \bibinfo {volume} {142}\ (\bibinfo  {publisher} {Elsevier},\ \bibinfo {year} {2004})\BibitemShut {NoStop}%
\bibitem [{\citenamefont {d’Ariano}\ \emph {et~al.}(2004)\citenamefont {d’Ariano}, \citenamefont {Perinotti},\ and\ \citenamefont {Sacchi}}]{d2004informationally}%
  \BibitemOpen
  \bibfield  {author} {\bibinfo {author} {\bibfnamefont {G.}~\bibnamefont {d’Ariano}}, \bibinfo {author} {\bibfnamefont {P.}~\bibnamefont {Perinotti}},\ and\ \bibinfo {author} {\bibfnamefont {M.}~\bibnamefont {Sacchi}},\ }\bibfield  {title} {\bibinfo {title} {Informationally complete measurements and group representation},\ }\href@noop {} {\bibfield  {journal} {\bibinfo  {journal} {Journal of Optics B: Quantum and Semiclassical Optics}\ }\textbf {\bibinfo {volume} {6}},\ \bibinfo {pages} {S487} (\bibinfo {year} {2004})}\BibitemShut {NoStop}%
\bibitem [{\citenamefont {Garc{\'\i}a-P{\'e}rez}\ \emph {et~al.}(2021)\citenamefont {Garc{\'\i}a-P{\'e}rez}, \citenamefont {Rossi}, \citenamefont {Sokolov}, \citenamefont {Tacchino}, \citenamefont {Barkoutsos}, \citenamefont {Mazzola}, \citenamefont {Tavernelli},\ and\ \citenamefont {Maniscalco}}]{garcia2021learning}%
  \BibitemOpen
  \bibfield  {author} {\bibinfo {author} {\bibfnamefont {G.}~\bibnamefont {Garc{\'\i}a-P{\'e}rez}}, \bibinfo {author} {\bibfnamefont {M.~A.}\ \bibnamefont {Rossi}}, \bibinfo {author} {\bibfnamefont {B.}~\bibnamefont {Sokolov}}, \bibinfo {author} {\bibfnamefont {F.}~\bibnamefont {Tacchino}}, \bibinfo {author} {\bibfnamefont {P.~K.}\ \bibnamefont {Barkoutsos}}, \bibinfo {author} {\bibfnamefont {G.}~\bibnamefont {Mazzola}}, \bibinfo {author} {\bibfnamefont {I.}~\bibnamefont {Tavernelli}},\ and\ \bibinfo {author} {\bibfnamefont {S.}~\bibnamefont {Maniscalco}},\ }\bibfield  {title} {\bibinfo {title} {Learning to measure: Adaptive informationally complete generalized measurements for quantum algorithms},\ }\href@noop {} {\bibfield  {journal} {\bibinfo  {journal} {Prx quantum}\ }\textbf {\bibinfo {volume} {2}},\ \bibinfo {pages} {040342} (\bibinfo {year} {2021})}\BibitemShut {NoStop}%
\bibitem [{\citenamefont {Renes}\ \emph {et~al.}(2004)\citenamefont {Renes}, \citenamefont {Blume-Kohout}, \citenamefont {Scott},\ and\ \citenamefont {Caves}}]{renes2004symmetric}%
  \BibitemOpen
  \bibfield  {author} {\bibinfo {author} {\bibfnamefont {J.~M.}\ \bibnamefont {Renes}}, \bibinfo {author} {\bibfnamefont {R.}~\bibnamefont {Blume-Kohout}}, \bibinfo {author} {\bibfnamefont {A.~J.}\ \bibnamefont {Scott}},\ and\ \bibinfo {author} {\bibfnamefont {C.~M.}\ \bibnamefont {Caves}},\ }\bibfield  {title} {\bibinfo {title} {Symmetric informationally complete quantum measurements},\ }\href@noop {} {\bibfield  {journal} {\bibinfo  {journal} {Journal of Mathematical Physics}\ }\textbf {\bibinfo {volume} {45}},\ \bibinfo {pages} {2171} (\bibinfo {year} {2004})}\BibitemShut {NoStop}%
\bibitem [{\citenamefont {Stricker}\ \emph {et~al.}(2022)\citenamefont {Stricker}, \citenamefont {Meth}, \citenamefont {Postler}, \citenamefont {Edmunds}, \citenamefont {Ferrie}, \citenamefont {Blatt}, \citenamefont {Schindler}, \citenamefont {Monz}, \citenamefont {Kueng},\ and\ \citenamefont {Ringbauer}}]{stricker2022experimental}%
  \BibitemOpen
  \bibfield  {author} {\bibinfo {author} {\bibfnamefont {R.}~\bibnamefont {Stricker}}, \bibinfo {author} {\bibfnamefont {M.}~\bibnamefont {Meth}}, \bibinfo {author} {\bibfnamefont {L.}~\bibnamefont {Postler}}, \bibinfo {author} {\bibfnamefont {C.}~\bibnamefont {Edmunds}}, \bibinfo {author} {\bibfnamefont {C.}~\bibnamefont {Ferrie}}, \bibinfo {author} {\bibfnamefont {R.}~\bibnamefont {Blatt}}, \bibinfo {author} {\bibfnamefont {P.}~\bibnamefont {Schindler}}, \bibinfo {author} {\bibfnamefont {T.}~\bibnamefont {Monz}}, \bibinfo {author} {\bibfnamefont {R.}~\bibnamefont {Kueng}},\ and\ \bibinfo {author} {\bibfnamefont {M.}~\bibnamefont {Ringbauer}},\ }\bibfield  {title} {\bibinfo {title} {Experimental single-setting quantum state tomography},\ }\href@noop {} {\bibfield  {journal} {\bibinfo  {journal} {PRX Quantum}\ }\textbf {\bibinfo {volume} {3}},\ \bibinfo {pages} {040310} (\bibinfo {year} {2022})}\BibitemShut {NoStop}%
\bibitem [{\citenamefont {Gour}\ and\ \citenamefont {Kalev}(2014)}]{gour2014construction}%
  \BibitemOpen
  \bibfield  {author} {\bibinfo {author} {\bibfnamefont {G.}~\bibnamefont {Gour}}\ and\ \bibinfo {author} {\bibfnamefont {A.}~\bibnamefont {Kalev}},\ }\bibfield  {title} {\bibinfo {title} {Construction of all general symmetric informationally complete measurements},\ }\href@noop {} {\bibfield  {journal} {\bibinfo  {journal} {Journal of Physics A: Mathematical and Theoretical}\ }\textbf {\bibinfo {volume} {47}},\ \bibinfo {pages} {335302} (\bibinfo {year} {2014})}\BibitemShut {NoStop}%
\bibitem [{\citenamefont {Tao}\ and\ \citenamefont {Vu}(2011)}]{tao2011random}%
  \BibitemOpen
  \bibfield  {author} {\bibinfo {author} {\bibfnamefont {T.}~\bibnamefont {Tao}}\ and\ \bibinfo {author} {\bibfnamefont {V.}~\bibnamefont {Vu}},\ }\bibfield  {title} {\bibinfo {title} {{Random matrices: Universality of local eigenvalue statistics}},\ }\href {https://doi.org/10.1007/s11511-011-0061-3} {\bibfield  {journal} {\bibinfo  {journal} {Acta Mathematica}\ }\textbf {\bibinfo {volume} {206}},\ \bibinfo {pages} {127 } (\bibinfo {year} {2011})}\BibitemShut {NoStop}%
\bibitem [{\citenamefont {Beatty}(2025)}]{beatty2025wasserstein}%
  \BibitemOpen
  \bibfield  {author} {\bibinfo {author} {\bibfnamefont {E.}~\bibnamefont {Beatty}},\ }\bibfield  {title} {\bibinfo {title} {Wasserstein distances on quantum structures: an overview},\ }\bibfield  {journal} {\bibinfo  {journal} {arXiv preprint arXiv:2506.09794}\ }\href {https://doi.org/10.48550/arXiv.2506.09794} {10.48550/arXiv.2506.09794} (\bibinfo {year} {2025}),\ \bibinfo {note} {41 pages},\ \Eprint {https://arxiv.org/abs/2506.09794} {arXiv:2506.09794 [quant-ph]} \BibitemShut {NoStop}%
\bibitem [{\citenamefont {Verteletskyi}\ \emph {et~al.}(2020)\citenamefont {Verteletskyi}, \citenamefont {Yen},\ and\ \citenamefont {Izmaylov}}]{verteletskyi2020measurement}%
  \BibitemOpen
  \bibfield  {author} {\bibinfo {author} {\bibfnamefont {V.}~\bibnamefont {Verteletskyi}}, \bibinfo {author} {\bibfnamefont {T.-C.}\ \bibnamefont {Yen}},\ and\ \bibinfo {author} {\bibfnamefont {A.~F.}\ \bibnamefont {Izmaylov}},\ }\bibfield  {title} {\bibinfo {title} {Measurement optimization in the variational quantum eigensolver using a minimum clique cover},\ }\href@noop {} {\bibfield  {journal} {\bibinfo  {journal} {The Journal of chemical physics}\ }\textbf {\bibinfo {volume} {152}} (\bibinfo {year} {2020})}\BibitemShut {NoStop}%
\bibitem [{\citenamefont {Yen}\ \emph {et~al.}(2020)\citenamefont {Yen}, \citenamefont {Verteletskyi},\ and\ \citenamefont {Izmaylov}}]{yen2020measuring}%
  \BibitemOpen
  \bibfield  {author} {\bibinfo {author} {\bibfnamefont {T.-C.}\ \bibnamefont {Yen}}, \bibinfo {author} {\bibfnamefont {V.}~\bibnamefont {Verteletskyi}},\ and\ \bibinfo {author} {\bibfnamefont {A.~F.}\ \bibnamefont {Izmaylov}},\ }\bibfield  {title} {\bibinfo {title} {Measuring all compatible operators in one series of single-qubit measurements using unitary transformations},\ }\href@noop {} {\bibfield  {journal} {\bibinfo  {journal} {Journal of chemical theory and computation}\ }\textbf {\bibinfo {volume} {16}},\ \bibinfo {pages} {2400} (\bibinfo {year} {2020})}\BibitemShut {NoStop}%
\bibitem [{\citenamefont {Chen}\ and\ \citenamefont {Lawrence}(1998)}]{chen1998linear}%
  \BibitemOpen
  \bibfield  {author} {\bibinfo {author} {\bibfnamefont {Y.}~\bibnamefont {Chen}}\ and\ \bibinfo {author} {\bibfnamefont {N.}~\bibnamefont {Lawrence}},\ }\bibfield  {title} {\bibinfo {title} {On the linear statistics of hermitian random matrices},\ }\href@noop {} {\bibfield  {journal} {\bibinfo  {journal} {Journal of Physics A: Mathematical and General}\ }\textbf {\bibinfo {volume} {31}},\ \bibinfo {pages} {1141} (\bibinfo {year} {1998})}\BibitemShut {NoStop}%
\bibitem [{\citenamefont {Holevo}(2024)}]{holevo2024estimates}%
  \BibitemOpen
  \bibfield  {author} {\bibinfo {author} {\bibfnamefont {A.}~\bibnamefont {Holevo}},\ }\bibfield  {title} {\bibinfo {title} {On estimates of trace-norm distance between quantum gaussian states},\ }\href@noop {} {\bibfield  {journal} {\bibinfo  {journal} {arXiv preprint arXiv:2408.11400}\ } (\bibinfo {year} {2024})}\BibitemShut {NoStop}%
\bibitem [{\citenamefont {Zhang}\ and\ \citenamefont {Rajabpour}(2023)}]{zhang2023trace}%
  \BibitemOpen
  \bibfield  {author} {\bibinfo {author} {\bibfnamefont {J.}~\bibnamefont {Zhang}}\ and\ \bibinfo {author} {\bibfnamefont {M.}~\bibnamefont {Rajabpour}},\ }\bibfield  {title} {\bibinfo {title} {Trace distance between fermionic gaussian states from a truncation method},\ }\href@noop {} {\bibfield  {journal} {\bibinfo  {journal} {Physical Review A}\ }\textbf {\bibinfo {volume} {108}},\ \bibinfo {pages} {022414} (\bibinfo {year} {2023})}\BibitemShut {NoStop}%
\end{thebibliography}%

\appendix

\section{Calculating covariance matrix for the correlated tomographic protocol}
\label{sec:app-sigma-calc}

Consider Pauli strings $P_i$, $P_j$. We want to find $\Sigma_{ij} = \mathrm{Cov}[y_i, y_j]$, where the tomographic protocol constructs $\Delta\hat\rho = \mathcal{A}_E(\vec y) = \sum_j y_j P_j$. Let $M_i$ denote the number of experiments where $P_i$ is measured and $P_j$ is \textit{not}. Define $M_j$ similarly. Let $M_{ij}$ denote the number of experiments in which both $P_i$ and $P_j$ are measured. So if $M^\mathrm{tot}_i = M_i + M_{ij}$, $M^\mathrm{tot}_j = M_j + M_{ij}$ denote the total number of experiments in which $P_i$ and $P_j$ are measured respectively, then since $\mathbb{E}[y_i] = \mathbb{E}[y_j] = 0$,
\begin{widetext}
\begin{equation}
\label{eq:covariance}
    \mathrm{Cov}(y_i, y_j) = \mathbb{E}[y_iy_j] = \frac{1}{M^\mathrm{tot}_i M^\mathrm{tot}_j }\mathbb{E}\left[\left(\sum_{k=1}^{M_i}Y^{(\mathrm{indep})}_{i, k} + \sum_{k=1}^{M_{ij}}Y^{(\mathrm{simul})}_{i, k}\right)\left(\sum_{k=1}^{M_j}Y^{(\mathrm{indep})}_{j, k} + \sum_{k=1}^{M_{ij}}Y^{(\mathrm{simul})}_{j, k}\right)\right],
\end{equation}
\end{widetext}
where $Y^{(\mathrm{indep})}_{i, k}$ is the outcome of the $k\mathrm{th}$ measurement of $P_i - \mathrm{Tr}(\rho P_i)$, in the experiments that \textit{do not} also measure $P_j$, while  $Y^{(\mathrm{simul})}_{i, k}$ is the outcome of the $k\mathrm{th}$ measurement of $P_i - \mathrm{Tr}(\rho P_i)$, in the experiments that \textit{do} also measure $P_j$. Similarly for the $Y_{j, k}$ terms. Expanding out the terms being multiplied, once we take the expected value, everything will go to zero except for the cross terms that are both part of the simultaneous measurements. In other words, we can simplify \cref{eq:covariance} to
\begin{widetext}
\begin{align}
    \frac{1}{M^\mathrm{tot}_i M^\mathrm{tot}_j }\mathbb{E}\left[\sum_{k, k'}Y^{(\mathrm{simul})}_{i, k}Y^{(\mathrm{simul})}_{j, k'}\right] &= \frac{1}{M^\mathrm{tot}_i M^\mathrm{tot}_j }\sum_{k, k'}\mathbb{E}\left[Y^{(\mathrm{simul})}_{i, k}Y^{(\mathrm{simul})}_{j, k'}\right] \\ &= \frac{1}{M^\mathrm{tot}_i M^\mathrm{tot}_j }\sum_{k, k'}\delta_{k,k'}\left[\mathrm{Tr(\rho P_i P_j) - \mathrm{Tr}(\rho P_i)\mathrm{Tr}(\rho P_j)}\right] \\ &= \frac{M_{ij}}{M^\mathrm{tot}_i M^\mathrm{tot}_j }\left[\mathrm{Tr(\rho P_i P_j) - \mathrm{Tr}(\rho P_i)\mathrm{Tr}(\rho P_j)}\right].
\end{align}
\end{widetext}
Since $M^\mathrm{tot}_i = S3^{N-w(P_i)}$, $M^\mathrm{tot}_j = S3^{N-w(P_j)}$, and $M_{ij} = \delta^E_{ij}S3^{N-w(P_i, P_j)}$, we find the results of \cref{eq:sophisticated-covariance} and \cref{eq:sophisticated-covariance-ratio}.

\end{document}